\documentclass[runningheads,dvipsnames]{llncs}
\usepackage
{todonotes}
\usepackage{changes}
\definechangesauthor[name=PLG, color=orange]{PLG}
\definechangesauthor[name=Mazen, color=red]{MF}
\definechangesauthor[name=Elias, color=blue]{EK}
\usepackage{comment}
\usepackage{graphicx}
\usepackage{amsmath}
\usepackage{hyperref}
\usepackage{wrapfig}
\usepackage[numbers]{natbib}
\usepackage[misc,geometry]{ifsym}
\usepackage{amsfonts}

\usepackage{listings}
\definecolor{cdarkgreen}{rgb}{0.0,0.4,0.0}
\definecolor{customblue}{rgb}{0.0,0.0,0.7}
\definecolor{cbluegreen}{rgb}{0.0,0.4,0.7}
\definecolor{cpurple}{rgb}{0.5,0.0,0.7}
\definecolor{corange}{rgb}{0.8,0.6,0.2}
\definecolor{cgreen}{rgb}{0,0.6,0}
\definecolor{OliveGreen}{RGB}{85,107,47}
\definecolor{Purple}{rgb}{0.5,0.0,0.7}
\colorlet{commentcolour}{green!50!black}
\colorlet{keywordcolour}{magenta!90!black}
\definecolor{MyDarkGreen}{rgb}{0.0,0.4,0.0}
\definecolor{MyBlue}{rgb}{0.0,0.0,0.7}
\definecolor{MyPurple}{rgb}{0.7,0.0,0.7}
\definecolor{BlueGreen}{rgb}{13,152,186}
\usepackage[framemethod=TikZ]{mdframed}
\lstdefinestyle{acslstyle}{language=C++,                
  alsoletter={<==>,==>,\&\&,||,\\,->},
  keywords = [2]{assumes,requires,ensures,behavior,logic,assigns,axiomatic,type,predicate,axiom,<==>,==>,\&\&,||,assert,loop, invariant, variant, ghost, for},
  keywords = [3]{\\valid,\\separated,\\forall,\\result,\\exact,\\model,\\total_error,\\error,\\abs,\\cos,\\sin,\\set_model,\\sqrt},
  keywords = [4]{PROOF_TACTIC},
  keywordstyle=[1]\color{blue},
  keywordstyle=[2]\color{Purple},
  keywordstyle=[3]\color{cbluegreen},
  keywordstyle=[4]\color{BurntOrange},
  morekeywords=[2]{real}, 
  morecomment=[l][\color{OliveGreen}]{//},
  morecomment=[s][\color{OliveGreen}]{/*}{*/},
  moredelim=**[s][\color{OliveGreen}]{/*@}{*/},
  moredelim=**[l][\color{OliveGreen}]{//@},
  rulecolor=\color{black},
  alsoother = ,}

\lstdefinestyle{pvsstyle}{morecomment=[l]{\%},
    commentstyle=\color{Maroon},
    keywords = [2]{lambda,type,forall,conversion,var,if, then, else,
      endif, importing, begin, theory, end,
      auto_rewrite,lemma,theorem,macro,let,implies,and,in,table,endtable,axiom,iff,true,false},
    sensitive=false,
    keywordstyle = [2]\color{Purple},
    literate=*
    {@}{{{\color{blue}+}}}1
    {+}{{{\color{blue}+}}}1
    {*}{{{\color{blue}*}}}1
    {=}{{{\color{blue}=}}}1
    {:=}{{{\color{blue}:=}}}2
    {/}{{{\color{blue}/}}}1
    {-}{{{\color{blue}-}}}1
    {<}{{{\color{blue}<}}}1
    {>}{{{\color{blue}>}}}1
    {<=}{{{\color{blue}<=}}}2
    {>=}{{{\color{blue}>=}}}2
    {->}{{{\color{black}->}}}2}

\lstdefinestyle{lustrestyle}
    { keywords=[1]{imported, node, 
        bool, int, float,
        initial, let, tel, until, unless, type, var, when, whennot,
        match, if, then, else, state, do, done, resume, restart, returns, merge,
        pre, current, last, map, red, fill, default,
        fby, automaton, tail, implies, pre, assert,
        PROPERTY, or, and, requires, ensures, observer},
      morecomment=[l][\color{blue}]{--},
      morecomment=[s][\color{blue}]{(*}{*)},
      morecomment=[l][\color{blue}]{assert}
    }[keywords,comments]

\lstdefinestyle{amsl}{language=Matlab,
breakatwhitespace=false,    
breaklines=true, 
postbreak=\raisebox{0ex}[0ex][0ex]{\ensuremath{\color{red}\hookrightarrow\space}}, 
rulecolor=\color{black!40},
emphstyle=\color{blue},
keywordstyle=\color{keywordcolour}\bfseries,
commentstyle=\color{commentcolour}\slshape,
morekeywords = {length,lsqr,norm,transpose,trace}, 
keywords = [2]{assumes,requires,ensures,behavior,predicate,logic,ghost,type,\&\&,||,loop, invariant}, 
keywords = [3]{PROOF_TACTIC},
keywords = [4]{on, off}, 
keywords = [5]{vecs,mats,smat,krons,is_symmetric,is_matrix,is_vector,type}, 
keywordstyle=[1]\color{MyBlue}, 
keywordstyle=[2]\color{cpurple},
keywordstyle=[3]\color{corange},
keywordstyle=[4]\color{MyPurple}, 
keywordstyle=[5]\color{cbluegreen}, 
morecomment=[l][\color{cgreen}]{//},
morecomment=[s][\color{cgreen}]{/*}{*/},
moredelim=**[s][\color{cgreen}]{/*@}{*/},
moredelim=**[l][\color{cgreen}]{//@},
}
\lstdefinestyle{matl}{language=Matlab,
breakatwhitespace=false,    
breaklines=true, 
rulecolor=\color{black!40},
emphstyle=\color{blue},
keywordstyle=\color{keywordcolour}\bfseries,
commentstyle=\color{commentcolour}\slshape,
morekeywords = {length,lsqr,norm,transpose,trace}, 
keywords = [2]{assumes,requires,ensures,behavior,predicate,logic,ghost,type,\&\&,||,loop, invariant}, 
keywords = [3]{PROOF_TACTIC},
keywords = [4]{on, off}, 
keywords = [5]{vecs,mats,smat,krons,is_symmetric,is_matrix,is_vector,type}, 
keywordstyle=[1]\color{MyBlue}, 
keywordstyle=[2]\color{cpurple},
keywordstyle=[3]\color{corange},
keywordstyle=[4]\color{MyPurple}, 
keywordstyle=[5]\color{cbluegreen}, 
morecomment=[l][\color{cgreen}]{//},
morecomment=[s][\color{cgreen}]{/*}{*/},
moredelim=**[s][\color{cgreen}]{/*@}{*/},
moredelim=**[l][\color{cgreen}]{//@},
}

\makeatletter
\def\mdf@@codeheading{Code Listings}
\define@key{mdf}{title}{%
   \def\mdf@@codeheading{#1}
}
\mdfdefinestyle{lstlisting}{%
 innertopmargin=5pt,
 innerbottommargin=0pt,
 middlelinewidth=1pt,
 outerlinewidth=2pt,outerlinecolor=white,
 innerleftmargin=2pt,
 innerrightmargin=2pt,
 leftmargin=-9pt,rightmargin=-9pt,
 skipabove=\topskip,
 skipbelow=\topskip,
 roundcorner=5pt,
 singleextra={\node[draw, fill=white,anchor=east, rounded corners=5pt,
     xshift=-10pt] at (P) {\csname mdf@@codeheading\endcsname};},
 firstextra={\node[draw, fill=white,anchor=east, rounded corners=5pt,
     xshift=-10pt] at (P) {\csname mdf@@codeheading\endcsname};}
}
\makeatother
\lstnewenvironment{acsl}{%
  \lstset{style=acslstyle,mathescape}%
  \mdframed[style=lstlisting,title={ACSL}]%
}{\endmdframed}

\lstnewenvironment{cacsl}{%
  \lstset{style=acslstyle,mathescape}%
  \mdframed[style=lstlisting,title={C+ACSL}]%
}{\endmdframed}

\lstnewenvironment{ccode}{%
  \lstset{style=acslstyle,mathescape}%
 \mdframed[style=lstlisting,title={C},nobreak=true]%
 \vspace{-.8em}
}{\endmdframed}

\lstnewenvironment{pvs}{%
  \lstset{style=pvsstyle,mathescape}%
  \mdframed[style=lstlisting,title={PVS}]%
}{\endmdframed}

\lstnewenvironment{matlab}{%
  \lstset{style=amsl,mathescape}%
 \mdframed[style=lstlisting,title={Matlab},nobreak=true]%
 \vspace{-.8em}
}{\endmdframed}

\lstnewenvironment{ccodenum}{%
  \lstset{style=acslstyle,mathescape,numbers=left,xleftmargin=2em}%
  \mdframed[style=lstlisting,title={C},nobreak=true]%
 \vspace{-.8em}
}{\endmdframed}

\lstnewenvironment{pseudocode}[1][]
{\lstset{style=pseudocodestyle, #1}}{}

\definecolor{module}{cmyk}{0.4,1,0.3,0.5}
\definecolor{constructor}{cmyk}{0,0.9,0,0.3}
\definecolor{function}{rgb}{0.1,0.65,0.1}
\definecolor{type}{rgb}{0.8,0,0}
\definecolor{variable}{rgb}{0,0,1}

\lstnewenvironment{ocaml}{%
  \lstset{language=[Objective]Caml,
  alsoletter={\textunderscore},
  basicstyle=\ttfamily,
  keywordstyle=\bfseries,
  moredelim=[is][\color{variable}]{//}{//},
  moredelim=[is][\color{type}]{!!}{!!},
  moredelim=[is][\color{function}]{--}{--},
  emphstyle=[1]{\color{variable}},
  emphstyle=[2]{\color{module}},
  emphstyle=[3]{\color{constructor}},
  emphstyle=[4]{\color{function}},
  emphstyle=[5]{\color{type}},
  identifierstyle=\color{variable},
  commentstyle=\itshape\color{gray},
 basewidth=0.45em,
  emph=[3]{None,Some},
  emph=[5]{float,int,bool,ref,unit,option},
}%
  \mdframed[style=lstlisting,title={OCaml},,nobreak=true]%
}{\endmdframed}

\lstdefinestyle{pseudocodestyle}{
    basicstyle=\ttfamily\small,
    mathescape=true,
    frame=single,
    framesep=5pt,
    breaklines=true,
    postbreak=\mbox{\textcolor{gray}{$\hookrightarrow$}\space},
    showstringspaces=false,
    numbers=none,
    captionpos=b,
    morekeywords={*,...},
    keywordstyle=\color{blue},
    commentstyle=\color{green},
    stringstyle=\color{red}
}

\newcommand{\fl}[2][]{{\rm fl}_{#1}\hspace{-0.1em}\left(#2\right)}

\newcommand{\pwiqc}{{\rm{pwIQC}}}
\newcommand{\bmat}[1]{\begin{bmatrix}#1\end{bmatrix}}

\newcommand{\iqc}{{\rm{IQC}}}

\newcommand{\innerproduct}[2]{\left\langle #1, #2 \right\rangle}

\begin{document}

\title{Formally Proving Invariant Systemic Properties of Control Programs Using Ghost Code and Integral Quadratic Constraints\thanks{This material is based upon work supported by the Army Research Office (ARO) under Contract No. W911NF-21-1-0250.}}
\titlerunning{Formally Proving Systemic Properties Using Ghost Code and IQCs}

 \author{Elias Khalife\inst{1}\textsuperscript{(\Letter)} \and
 Pierre-Loic Garoche\inst{2} \and
 Mazen Farhood\inst{1}}

\authorrunning{E. Khalife et al.}

 \institute{Virginia Tech \\ Kevin T. Crofton Department of Aerospace and Ocean Engineering
 \email{\{eliask,farhood\}@vt.edu}\\
 \and F\'{e}d\'{e}ration ENAC ISAE-SUPAERO ONERA, Universit\'{e} de Toulouse, France \\
 \email{pierre-loic.garoche@enac.fr}}
\maketitle
\begin{abstract}
This paper focuses on formally verifying invariant properties of control programs both at the model and code levels. The physical process is described by an uncertain discrete-time state-space system, where the dependence of the state-space matrix-valued functions defining the system on the uncertainties can be rational. The proposed approaches make use of pointwise integral quadratic constraints (IQCs) to characterize the uncertainties affecting the behavior of the system. Various uncertainties can be characterized by pointwise IQCs, including static linear time-varying perturbations and sector-bounded nonlinearities. Using the IQC framework, a sound overapproximation of the uncertain system, which is expressible at the code level, is constructed. Tools such as Frama-C, ACSL, WP, and an Alt-Ergo plugin are employed to ensure the validity of the state and output invariant properties across both real and float models. The first proposed approach can be used to formally verify (local) invariant properties of the control code. This capability is demonstrated in a couple of examples involving gain-scheduled path-following controllers designed for an uncrewed aircraft system and an autonomous underwater vehicle. The second  approach enables the verification of closed-loop invariant properties, i.e., invariant properties of the controlled system as a whole, in both real and float models, while preserving the integrity of the executable controller code. This is achieved by using ghost code attached to the control code for all elements related to the plant model with uncertainties, as the ghost code does not interfere with the executable code. The effectiveness of this approach is demonstrated in two examples on the control of a four-thruster hovercraft and the control of a two-mass rotational system.
\end{abstract}


\section{Introduction}\label{sec:Introduction}
The formal verification of the properties of a dynamical system is a challenging task, especially when faced with uncertainties or multiple thereof affecting its behavior \cite{peled2013}. Generally, deriving an exact mathematical abstraction of a physical process is not feasible due to inevitable uncertainties, often arising from intentional or unavoidable undermodeling. Moreover, computer-driven systems require discretization for the practical implementation of algorithms on physical processes. Therefore, it is reasonable to represent a controlled physical process as an uncertain, discrete-time system. This paper deals with the formal verification of invariant properties of such uncertain systems. Specifically, we aim to formally prove not only local properties of the control code, that is, properties related to the state and output variables of the controller, but also systemic properties that pertain to the state variables of the entire closed-loop system.

 In control systems theory, an invariant set is defined as a subset of the state space such that if the system's state ever enters this set, it remains there at all subsequent times \cite{blanchini1999set}. Ellipsoidal invariant sets are appealing because of their computability via linear matrix inequality  techniques  \cite{boyd1994linear} and sum-of-squares programming \cite{TOPCU2008}. They also feature implicitly in Lyapunov's stability theory when the storage function is a strictly convex quadratic function \cite{Khalil}. In this paper, we adopt the integral quadratic constraint (IQC) framework \cite{megretski1997,veenman2016} to conduct analysis on the uncertain systems of interest and ultimately determine invariant ellipsoidal sets. IQC analysis generally requires expressing the uncertain system in a linear fractional transformation (LFT) form. Thus,  we focus on uncertain state-space systems, where the state-space matrix-valued functions have at most rational dependence on the uncertainties. These uncertain systems can be expressed in the LFT framework as the interconnection of a linear time-invariant (LTI) nominal system and a structured uncertainty operator. In robust control theory, control analysis and synthesis problems involving LFT systems are well-studied; see, for instance, \cite{zhou1996robust,Zhou96,PACKARD199479}. We model an uncertain system, in which the uncertainties can be characterized with pointwise IQCs, using an augmented LTI system coupled with a quadratic constraint. This sound overapproximation of the uncertain system is expressible at the code level. Various  uncertainty sets, such as those of static linear time-varying (LTV) perturbations, sector-bounded nonlinearities,  time delays, and combinations thereof, admit pointwise IQC characterizations \cite{invariantpwIQC}. We then apply  results from \cite{invariantpwIQC} to obtain state and output bounding ellipsoids for that augmented system at the model level. These ellipsoids are used to express the system's invariant properties via first-order logic predicates,   amenable to Dijkstra's weakest precondition algorithm \cite{dijkstra1975guarded}, and the resultant proof objectives can be discharged~by~SMT~solvers.

 The uncertain systems described in the preceding could correspond to a closed-loop system, i.e., the interconnection of a dynamical system and a feedback controller, or just the controller, e.g., a stable linear parameter-varying (LPV) controller. We demonstrate in this work how to exploit results and concepts from IQC theory to formally verify invariant properties, obtained by applying IQC-based analysis results, in both real and float models using source-code analysis tools such as Frama-C \cite{Frama-C}. For instance, in two of the provided examples, we formally verify (local) invariant properties of field-tested path-following LPV controllers for a fixed-wing uncrewed aircraft system (UAS) and an autonomous underwater vehicle (AUV). As the uncertain systems of interest can have rational dependence on combinations of uncertainties, including LTV perturbations, sector-bounded nonlinearities, and uncertain time delays, and hence are challenging to express at code level, the formal verification will be carried out on a sound overapproximation of the uncertain system obtained from applying the IQC-based analysis approach. The overapproximation is in the form of an augmented LTI system, along with a quadratic constraint.
 In addition, in this work we devise an approach for formally verifying systemic properties of unmodified control code while accounting for floating-point computation errors, which is in line with regulatory requirements such as DO-178C regulations for airborne systems \cite{rtca2011airborne}. This approach applies to LTI controllers, where the dynamical system modeling the physical process can be any of the uncertain systems under consideration and will be expressed at code level as described before using  ghost code. An example on the control of a four-thruster hovercraft is also provided.  Furthermore, an extension to this approach is given to deal with affine LPV controllers, along with an illustrative example on the control of a two-mass rotational system.

\subsubsection*{Related Work.}
Closely related to our work, \cite{feron2010control} establishes axiomatic semantics of system properties at the model level. That work is further extended in \cite{wang2016formal}, where system properties are verified at the code level; however, the verification is limited to simpler systems with few states. The work \cite{DBLP:conf/hybrid/RouxJG15} addresses closed-loop system verification by encoding the entire closed-loop system in code, which does not verify the actual executable controller code. In contrast, our approach keeps the original controller code intact while associating it with a stateful contract that describes the closed-loop uncertain system using ghost code. Ghost code allows attaching specifications to the code in comments, that is, without modifying the program source instructions.
Moreover, the methods in \cite{wang2016formal} and \cite{DBLP:conf/hybrid/RouxJG15} do not tackle uncertain systems. The work in \cite{NFM2023} adopts a similar approach to the one presented in this paper, focusing on verifying numerical properties at the code level, including considerations for floating-point computation errors. However, there are several limitations in that work which have been addressed in this paper. One limitation is  the approach in \cite{NFM2023} is restricted to LPV systems with affine parameter dependence, i.e., uncertain systems whose state-space matrix-valued functions depend affinely on static linear time-varying perturbations, which are expressible at the code level. Also, the tractability of the approach in \cite{NFM2023} is constrained by the size of the system and the number of parameters, as it does not scale well with increases in the number of state variables and parameters. In contrast, the approach  in this paper can handle relatively large-sized, complex uncertain systems typically not expressible at the code level, such as those with rational dependence on the parameters. Last, as with \cite{DBLP:conf/hybrid/RouxJG15}, \cite{NFM2023} cannot express and analyze closed-loop properties of the controller code without modifying it.

\subsubsection*{Contributions.}
The following work makes several contributions. It
\begin{itemize}
\item Integrates the IQC framework into deductive verification, enabling complex uncertain systems to be modeled, expressed, and verified at the code level.
\item Presents a method for formally verifying state and output invariant properties of uncertain systems in  real and float models, addressing floating-point error considerations by modifying Hoare triples to account for these~errors.
\item Gives a method to verify  properties of closed-loop systems comprising an LTI controller and an uncertain plant, using ghost code to capture plant dynamics without changing the  controller code, with verification conducted in both real and float models and an extension proposed for affine~LPV~controllers.
\item Demonstrates the applicability of the proposed approaches using tools such as Frama-C \cite{Frama-C}, with adaptability to other environments like SPARK/Ada~\cite{barnes1984programming}, and provides annotation templates for implementation in the  Appendices \ref{sec:frama-c_template} and \ref{sec:frama-c-ghost}.
\item Gives examples showcasing the effectiveness of the IQC-based  formal verification approach, whereby local properties of field-tested path-following LPV controllers for a UAS and an AUV, with relatively large number of state, input, and output variables and polynomial dependence on the scheduling parameters, are formally verified  while accounting for floating-point errors.
\item Provides  two examples on formally verifying systemic properties of unmodified control code, whereby systemic properties of an LTI control program for a four-thruster hovercraft  and those of an affine LPV controller for a two-mass rotational system are formally verified in both real and float models.
\end{itemize}

\subsubsection*{Structure and Organization.}The paper is structured as follows. Section~\ref{sec:Preliminaries} introduces the uncertain system, explains the construction of the augmented system with IQCs that overapproximates the uncertain system, and provides existing results for finding state and output bounding ellipsoids. Section \ref{sec:VerificationOfInvariants} details the verification process of the local state and output invariant properties of the augmented system in both real and float models. In Section \ref{sec:closedloopverification}, we present the approach to formally verify systemic properties of a control program using ghost code. Section \ref{sec:examples} illustrates the applicability of the proposed verification methods in four examples. The paper concludes with Section \ref{sec:conclusion}. Appendix \ref{sec:LFT} describes how to formulate an uncertain system in the LFT framework. Appendices \ref{sec:frama-c_template} and \ref{sec:frama-c-ghost} provide templates for the application of the proposed approaches.

\subsubsection*{Notation.}
$\mathbb{R}^n$, $\mathbb{R}^{m \times n}$, and $\mathbb{S}^{n}_{++}$ ($\mathbb{S}^{n}_{+}$) denote the sets of real vectors of dimension $n$, $m\times n$ real matrices, and $n\times n$ real symmetric positive definite (positive semidefinite) matrices, respectively. Given $X\in\mathbb{S}^{n}_{++}$, we define the ellipsoid $\mathcal{E}_X=\left\{z\in\mathbb{R}^n\mid z^\top X z\leq1\right\}$. $X^*$ denotes the adjoint of $X$, and $Y^\top$ denotes the transpose of $Y$. The identity matrix and identity operator are both denoted by $I$. We write $X\succ0$ ($X\succeq0$) to indicate that $X$ is positive definite (positive semidefinite). The block-diagonal augmentation of matrices (or operators) $X_1,\ldots,X_n$ is denoted by $\text{blkdiag}(X_1,\ldots,X_n)$. The Hilbert space of square-summable vector-valued sequences $x=(x(0),x(1),\ldots)$, with $x(k)\in\mathbb{R}^n$,  is denoted by $\ell_{2}^n$; i.e., $\innerproduct{x}{x}_{\ell_2}=\sum_{k=0}^{\infty}x(k)^\top x(k)<\infty$, where $\innerproduct{\cdot}{\cdot}_{\ell_2}$ denotes the inner product on $\ell_2$ (the dimension $n$ may be suppressed when irrelevant or clear from  context).

\section{Preliminaries}\label{sec:Preliminaries}
\subsection{IQC Framework}\label{sec:IQC_Framework}
\paragraph{Uncertainty Sets.}
Consider the uncertainty set $\mathbf{\Delta}$, consisting of bounded causal uncertainty operators $\Delta$ on $\ell_2$.
 Suppose  $\vartheta = \Delta(\varphi)$, where $\varphi=(\varphi(0),\varphi(1),\ldots)\in\ell_2^{n_\varphi}$, $\vartheta=(\vartheta(0),\vartheta(1),\dots)\in\ell_2^{n_\vartheta}$, and $\Delta:\ell_2^{n_\varphi}\rightarrow\ell_2^{n_\vartheta}$.
The set $\mathcal{M}_{\mathbf{\Delta}}$, containing all signal pairs $(\varphi, \vartheta)$ satisfying the preceding equation for any $\Delta \in \mathbf{\Delta}$, is defined~as
    \[\mathcal{M}_{\mathbf{\Delta}} := \left\{ (\varphi, \vartheta) \mid \vartheta = \Delta(\varphi),\, \varphi \in \ell_2^{n_\varphi},\,\text{and} \ \Delta \in \mathbf{\Delta} \right\}.\]
Given a causal, bounded, linear operator $\Psi$ and a self-adjoint, memoryless operator $\bar{S}$, i.e., $\bar{S}=\text{blkdiag}(\bar{S}(0),\bar{S}(1),\ldots)=\bar{S}^*$, defining the so-called IQC multiplier $\Pi=\Psi^*\bar{S}\Psi$,  the set $\mathcal{M}_\Pi$ is defined as
\begin{equation}\label{Mpi}
    \mathcal{M}_\Pi := \Bigg\{ (\varphi, \vartheta)  \ \bigg|\ \varphi\in\ell_2^{n_\varphi},\,\innerproduct{\bmat{\varphi \\ \vartheta}}{\Pi\bmat{\varphi\\\vartheta}}_{\ell_2} \geq 0 \Bigg\}.
\end{equation}
We say the uncertainty set $\mathbf{\Delta}$ satisfies the IQC defined by $\Pi$,  or $\mathbf{\Delta}\in\iqc(\Pi)$, if $\mathcal{M}_{\mathbf{\Delta}} \subseteq \mathcal{M}_\Pi$ \cite{megretski1997, veenman2016, fry2021robustness}; i.e., the set defined by the IQC constitutes a sound overapproximation of the set $\mathcal{M}_{\mathbf{\Delta}}$, which characterizes the uncertainty set $\mathbf{\Delta}$.

\begin{figure}[t]
    \parbox{2.5in}{\centering\includegraphics[scale=0.35]{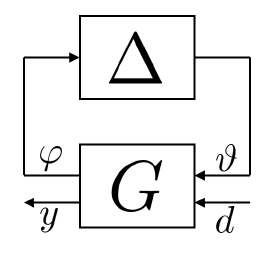}
    \caption{LFT system $(G,\Delta)$.}
    \label{fig:LFTsystem}}
    \parbox{2.25in}{\centering\includegraphics[scale=0.25]{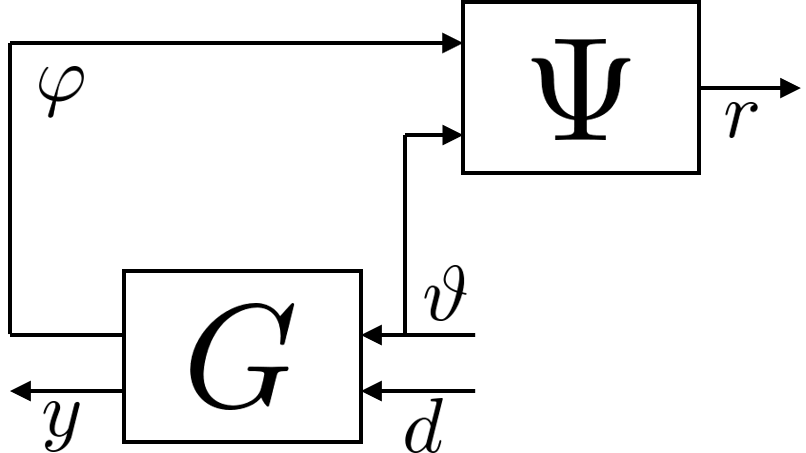}
    \caption{Augmented system $H$.}
    \label{fig:lftIQC}}
\end{figure}

\paragraph{IQCs.}
Suppose $\mathbf{\Delta}\in\iqc(\Pi)$, where $\Pi=\Psi^*\bar{S}\Psi$ and $\Psi$, $\bar{S}$ are as defined before. Although analysis results based on time-varying IQC multipliers have been developed in the literature \cite{fry2021robustness,farhood2024RNC}, we will focus the discussion here on time-invariant multipliers for simplicity. In this case, the linear, bounded, causal operator $\Psi$, referred to as the IQC filter,  can be represented by a stable LTI system with inputs $\varphi$ and $\vartheta$ and output  $r \in \ell_{2e}^{n_r}$ (see Fig.~\ref{fig:lftIQC}), defined by the  state-space equations
\begin{equation*}
\begin{aligned}
x_{\Psi}(k+1) &= A_{\Psi}x_{\Psi}(k) + B_{\Psi_1}\varphi(k) + B_{\Psi_2}\vartheta(k), \ \ x_{\Psi}(0) = 0\in \mathbb{R}^{n_{\Psi}},\\
r(k) &= C_{\Psi}x_{\Psi}(k) + D_{\Psi_1}\varphi(k) + D_{\Psi_2}\vartheta(k).
\end{aligned}\label{eq:iqcfilterequation}
\end{equation*}
The corresponding infinite-dimensional matrix representation for $\Psi$ would then~be
\begin{equation*}
    \Psi= \bmat{D_\Psi && 0 && 0 && \ldots\\
                C_\Psi B_\Psi && D_\Psi && 0 && \ldots\\
                C_\Psi A_\Psi B_\Psi && C_\Psi B_\Psi && D_\Psi && \ddots\\
                \vdots && \vdots && \vdots && \ddots}, \label{eq:IDM_Psi}
\end{equation*}
where  $B_\Psi=\bmat{B_{\Psi_1} && B_{\Psi_2}}$ and $D_\Psi=\bmat{D_{\Psi_1} && D_{\Psi_2}}$. The  memoryless operator $\bar{S}$  is defined in this case as $\bar{S}=\text{blkdiag}(S,S,\ldots)$, where  $S$ is an $n_r\times n_r$ symmetric matrix. Then, the IQC defining  $\mathcal{M}_\Pi$ in (\ref{Mpi}) can  be equivalently expressed~as
\begin{equation}
    \sum_{k=0}^{\infty} r(k)^\top S r(k) \geq 0.\label{eq:softIQC}
\end{equation}
The time-domain representation allows defining different types of IQCs. The condition in \ref{eq:softIQC} is called a soft IQC, and  numerous uncertainty sets admit soft IQC characterizations \cite{veenman2016}. Other types of IQCs include hard IQCs \cite{Hu2017}, finite-horizon IQCs with a terminal condition \cite{SCHERER2022150, scherer2022dissipativity}, and $\rho$-hard IQCs \cite{lessard2016analysis}. Of specific interest are pointwise IQCs, where  (\ref{eq:softIQC}) is replaced with the more restrictive condition
\begin{equation}
    r(k)^\top S r(k) \geq 0, \ \mbox{for all } k\geq 0. \label{eq:pwIQCcondition}
\end{equation}
Still, various uncertainties, including static LTI and LTV perturbations, uncertain time-varying delays, and sector-bounded nonlinearities,  admit pointwise IQC characterizations \cite{invariantpwIQC, koroglu2007, scherer2006, megretski1997, Schwenkel2022, veenman2016}. We write $\mathbf{\Delta}\in\pwiqc(\Psi,S)$ to indicate that $\mathbf{\Delta}$ satisfies the pointwise IQC defined by the factors $(\Psi,S)$.

In this work, we give examples on systems with parametric uncertainties. So, for illustrative purposes, consider the following set of static LTV uncertainties:
\begin{equation*}
    \mathbf{\Delta}_{\mathrm{SLTV}}:=\big\{\Delta \mid \vartheta=\Delta(\varphi) \equiv \big(\vartheta(k)=\delta(k)\varphi(k),\,
    |\delta(k)|\leq\alpha \text{ for all } k\geq0\big)\big\}.
\end{equation*}
We will now demonstrate how $\mathbf{\Delta}_{\mathrm{SLTV}}$ satisfies the pointwise IQC defined by $(\Psi_{\mathrm{SLTV}}, S_{\mathrm{SLTV}})$, where $\Psi_{\mathrm{SLTV}}=\text{blkdiag}(I,I)$ and the scaling matrix
\[
S_{\mathrm{SLTV}}=\bmat{\alpha^2 X & & Y \\ Y^\top & & -X}, \ \mbox{with} \ X\in\mathbb{S}^{n_{\varphi}}_+ \ \mbox{and} \ Y=-Y^\top.
\]
Specifically, for all $k\geq0$, $\varphi \in \ell_{2}^{n_{\varphi}}$, $\vartheta = \Delta(\varphi)$, and $\Delta\in\mathbf{\Delta}_{\mathrm{SLTV}}$, we have $r(k)=(\varphi(k),\vartheta(k))=(\varphi(k),\delta(k)\varphi(k))$ and $r(k)^\top S_{\mathrm{SLTV}}r(k)=(\alpha^2-\delta(k)^2)\varphi(k)^\top X\varphi(k)$. Thus, $r(k)^\top S_{\mathrm{SLTV}}r(k)\geq 0$  since $X\succeq0$ and $|\delta(k)|\leq \alpha$,  satisfying condition (\ref{eq:pwIQCcondition}).

\subsection{Uncertain System Modeling}
Consider an uncertain system $(G,\mathbf{\Delta})$ in LFT form (see Fig. \ref{fig:LFTsystem} and Appendix \ref{sec:LFT}), where $G$ is a stable discrete-time LTI system and $\Delta\in\mathbf{\Delta}$ is causal and bounded. We assume  $(G,\mathbf{\Delta})$ is well-posed, i.e., for every input signal and for each $\Delta \in \mathbf{\Delta}$, there exist unique signals that satisfy the system's state-space equations and causally depend on the input.
To derive a model for this uncertain system, we utilize the IQC framework to characterize the uncertainty set $\mathbf{\Delta}$. Specifically, assuming $\mathbf{\Delta}\in\pwiqc(\Psi,S)$, we construct an augmented LTI system $H$ from the nominal system $G$ and the IQC filter $\Psi$, as shown in Fig. \ref{fig:lftIQC}, which takes as inputs the signals $\vartheta$ and $d$~and~outputs~the~signals~$r$~and~$y$.

Clearly, the augmented system $H$ is an overapproximation of the uncertain system; however, without any additional constraints, this overapproximation is too conservative since the input signal $\vartheta$  can be any signal in $\ell_2^{n_{\vartheta}}$ and so is not constrained by the uncertainty ($\vartheta=\Delta(\varphi)$) as in the case of the uncertain system. By enforcing the IQC condition (\ref{eq:pwIQCcondition}) on the output signal $r$, the set of signal pairs $(\varphi,\vartheta)$ admitted in the augmented system shrinks but still encompasses the signal pairs admitted in the uncertain system, as discussed in Section \ref{sec:IQC_Framework}. The conservativeness of this overapproximation depends on how well the IQC multiplier characterizes the uncertainty. As the reachable states and outputs of the uncertain system $(G, \mathbf{\Delta})$ are also reachable by the augmented system $H$, coupled with the IQC condition (\ref{eq:pwIQCcondition}), this overapproximation  will be taken as a sound mathematical abstraction of the uncertain system in our verification approach. For all $k\geq0$, $H$ is described by the following state-space equations:
\begin{equation}
\begin{aligned}
x_H(k+1)&=A_Hx_H(k)+B_{H_1}\vartheta(k)+B_{H_2}d(k),\\
r(k)&=C_{H_1}x_H(k)+D_{H_{11}}\vartheta(k)+D_{H_{12}}d(k),\\
y(k)&=C_{H_2}x_H(k)+D_{H_{21}}\vartheta(k)+D_{H_{22}}d(k),
\end{aligned}\label{eq:augmentedsystem}
\end{equation}
with $x_H(k)=(x_G(k),x_{\Psi}(k))$, $C_{H_2}=\begin{bmatrix}C_{G_2} && 0\end{bmatrix}$, $D_{H_{21}}=D_{G_{21}}$, $D_{H_{22}}=D_{G_{22}}$,
\begin{align*}
A_H&=\begin{bmatrix}
A_G && 0\\
B_{\Psi_1}C_{G_1} && A_{\Psi}
\end{bmatrix}, & B_{H_1}&=\begin{bmatrix}
B_{G_1}\\
B_{\Psi_1}D_{G_{11}}+B_{\Psi_2}
\end{bmatrix}, &
B_{H_2}&=\begin{bmatrix}
B_{G_2}\\
B_{\Psi_1}D_{G_{12}}
\end{bmatrix}, \\ C_{H_1}&=\begin{bmatrix}
D_{\Psi_1}C_{G_1}&& C_{\Psi}
\end{bmatrix}, & D_{H_{11}}&=D_{\Psi_1}D_{G_{11}}+D_{\Psi_2}, &
 D_{H_{12}}&=D_{\Psi_1}D_{G_{12}}.
\end{align*}
If the IQC filter is static, i.e., $n_\Psi=0$, then $x_H=x_G$. However, it is crucial to recognize that, in this case, the states of the  system $H$ do not necessarily have the same physical meaning as the states of the uncertain system $(G,\mathbf{\Delta})$, and the reachable state sets of $H$ are supersets of those for the uncertain system.

\subsection{Preliminary Results}\label{sec:invariant_theorems}
Consider the augmented system $H$ defined in (\ref{eq:augmentedsystem}), coupled with condition \ref{eq:pwIQCcondition}), capturing the behavior of $(G,\mathbf{\Delta})$, where $\mathbf{\Delta} \in \pwiqc(\Psi, S)$. Assume that the exogenous input $d(k) \in \mathcal{D}$ for all $k \geq 0$, where $\mathcal{D}$ is a polytope with $t$ vertices $\hat{d}_1, \ldots, \hat{d}_t\in\mathbb{R}^{n_d}$ and includes the zero vector. We define the following functions:
\begin{align*}
     &\mathcal{F}_1(X_1, X_2, X_3) =
    \begin{bmatrix}
        -X_1^\top X_2 X_1 && -X_1^\top X_2 X_3\\
        -X_3^\top X_2 X_1 && 1 - X_3^\top X_2 X_3
    \end{bmatrix},\\
    &\mathcal{F}_2(X_1, X_2, X_3) =
    \begin{bmatrix}
        X_1^\top X_2 X_1 && X_1^\top X_2 X_3\\
        X_3^\top X_2 X_1 && X_3^\top X_2 X_3
    \end{bmatrix},\\
    &\mathcal{F}_3(X) = \text{blkdiag}(-X, 0_{n_\vartheta \times n_\vartheta}, 1),\\
    &\mathcal{F}_4(X_1, X_2) = X_1^\top X_2 X_1.
\end{align*}
The following results, provided in \cite{AboujaoudeGarocheFarhood21, invariantpwIQC}, give sufficient conditions for determining state-invariant and output-bounding ellipsoids for the system $H$.\vspace{2mm}
\begin{theorem}[\cite{AboujaoudeGarocheFarhood21}]\label{thm:state}
    If there exist $P\in\mathbb{S}_{++}^{n_H}$, $\tau_1\geq0$, and $\tau_2\geq0$ such that
    \[
        \mathcal{F}_1\left(\begin{bmatrix}A_H & B_{H_1}\end{bmatrix},P,B_{H_2}\hat{d}_i\right)-\tau_1\mathcal{F}_3\left(P\right)
        -\tau_2\mathcal{F}_4\left(S_x,\begin{bmatrix}C_{H_1} & D_{H_{11}} & D_{H_{12}}\end{bmatrix}\right)\succeq 0,
    \]
    for $i=1,\ldots,t$, and $B_{H_2}^\top P B_{H_2}+\tau_2 D_{H_{12}}^\top S_x D_{H_{12}}\succeq 0$,
    then $x_H(k)\in\mathcal{E}_P$ implies that $x_H(k+1)\in\mathcal{E}_P$ for any $k\geq 0$, i.e., $\mathcal{E}_P$ is a state-invariant ellipsoid for  $H$.
\end{theorem}
\vspace{2mm}
\begin{theorem}[\cite{invariantpwIQC}]\label{thm:output}
    Given $P\in\mathbb{S}_{++}^{n_H}$, if there exist $Q\in\mathbb{S}_{++}^{n_y}$, $\tau_3\geq0$, and $\tau_4\geq0$ such that, for $i=1,\ldots,t$,
    \begin{align*}
        \mathcal{F}_1(\begin{bmatrix}C_{H_2} & D_{H_{21}}\end{bmatrix},Q,D_{H_{22}}\hat{d}_i)-\tau_3\mathcal{F}_3(P)
        -\tau_4\mathcal{F}_2(\begin{bmatrix}C_{H_1} & D_{H_{11}}\end{bmatrix},S_\text{y},D_{H_{12}}\hat{d}_i)\succeq 0,
    \\
        D_{H_{22}}^\top Q D_{H_{22}}+\tau_4 D_{H_{12}}^\top S_\text{y} D_{H_{12}}\succeq 0,
    \end{align*}
    then $x_H(k)\in\mathcal{E}_P$ implies that $y(k)\in\mathcal{E}_Q$ for any $k\geq 0$, i.e., $\mathcal{E}_Q$ is an output-bounding ellipsoid for system $H$.
\end{theorem}

To find the state-invariant ellipsoid $\mathcal{E}_P$ for the system $H$ as per Theorem \ref{thm:state}, we solve a convex optimization problem by gridding $\tau_1$ over $[0,1]$ and absorbing $\tau_2$ into $S_x$. Upon obtaining $\mathcal{E}_P$, we solve another convex optimization problem based on the conditions in Theorem \ref{thm:output} to determine the output-bounding ellipsoid $\mathcal{E}_Q$, but this time absorbing $\tau_4$ into $S_y$. Both $S_x$ and $S_y$ are selected from the same set of scaling matrices, and both result in IQC multipliers that characterize the uncertainty set $\mathbf{\Delta}$; however, allowing $S_x$ and $S_y$ to differ reduces the conservatism of the bounding ellipsoids obtained. Additionally, \cite[Theorem 3]{NFM2023} enables the computation of state and output bounding ellipsoids for  $H$ when the exogenous input lies within an ellipsoid pointwise in time, i.e., $\mathcal{D} = \mathcal{E}_\Lambda$ for some $\Lambda \in \mathbb{S}_{++}^{n_d}$.

\section{Model-Level Verification of Invariants of Dynamical Systems with Floating-Point Arithmetic}\label{sec:VerificationOfInvariants}
This section presents the process of formally verifying state and output invariant properties of uncertain systems using Hoare logic as the foundational framework. While these properties are theoretically valid, they are not formally verified using deductive tools. We first outline the verification process in the real model, where the proofs are simpler due to mathematical operations with infinite precision. We then address the verification in the float model, highlighting the challenges posed by floating-point computations and our strategies to overcome them.
For a detailed application of these concepts through Frama-C, including practical implementations using its WP plugin and ACSL annotations, see~Appendix~\ref{sec:frama-c_template}.
\subsection{Expressing Invariant Properties in Hoare Logic}
In Hoare logic, the correctness of program execution is expressed through Hoare triples of the form $\{P\} \; c \; \{Q\}$, where $P$ and $Q$ are the precondition and postcondition, respectively, and $c$ is the command or sequence of commands. The semantics of $\{P\} \; c \; \{Q\}$ is that if $P$ holds before executing command $c$, then $Q$ will hold after $c$ has been executed, assuming $c$ terminates \cite{hoare1969axiomatic,hoare1976}.
In particular, an invariant $J$ in a loop is expressed in a Hoare triple as $\{J\} \; \text{body} \; \{J\}$, indicating that if $J$ holds at the beginning of the loop body, it will still hold at the end of each iteration of the loop body. This form ensures that the invariant $J$ is maintained throughout the execution of the loop. Note that the body of the loop can conceptually encompass the entire program, especially in control systems where the program runs in a \texttt{while true} loop.

\paragraph{State Invariant Property.}
The state invariant property of the augmented system $H$, defined in (\ref{eq:augmentedsystem}), which models the uncertain system $(G,\mathbf{\Delta})$, is expressed through a Hoare triple as follows:
\begin{gather*}
\{x_H(k) \in \mathcal{E}_P \land  d(k) \in \mathcal{D} \land r(k)^\top S_x r(k) \geq 0\} \;
\texttt{stateUpdate} \;
\{x_H(k+1) \in \mathcal{E}_P\},
\end{gather*}
where \texttt{stateUpdate} represents the function executing the state transition of system $H$ based on (\ref{eq:augmentedsystem}).
This formulation specifies that, provided $x_H(k)$ resides within $\mathcal{E}_P$, and given the constraints on $d(k)$ and $r(k)$ before the update, the state $x_H(k+1)$ post-update will also reside within $\mathcal{E}_P$, thus maintaining the state invariant. The preconditions $d(k) \in \mathcal{D}$ and $r(k)^\top S_x r(k) \geq 0$ are necessary for the state invariant to hold, as per Theorem \ref{thm:state}. In particular, the pointwise IQC condition $r(k)^\top S_x r(k) \geq 0$ is critical in ensuring that  $H$  captures the dynamics of the original uncertain system $(G,\mathbf{\Delta})$ without being overly conservative.

\paragraph{Output Invariant Property.}
Following the structure established for the state invariant, the output invariant property of the augmented system $H$, modeling $(G,\mathbf{\Delta})$, is captured by the following Hoare triple:
\begin{gather*}
\{x_H(k) \in \mathcal{E}_P \land d(k) \in \mathcal{D} \land r(k)^\top S_y r(k) \geq 0\}\;
\texttt{outputUpdate}\;
\{y(k) \in \mathcal{E}_Q\},
\end{gather*}
where \texttt{outputUpdate} denotes the function responsible for updating the output $y(k)$ of $H$ based on the system dynamics (\ref{eq:augmentedsystem}).

\subsection{Incremental Verification Process: From Real to Floating-Point Arithmetic Verification}\label{sec:verification_process}
Verification is conducted under two primary computational models: the real model and the float model.
In the real model, computations are assumed to occur with infinite precision, simplifying mathematical proofs by not accounting for rounding errors inherent in floating-point arithmetic. Verification of properties in the real model can be efficiently performed using automated theorem provers and SMT solvers such as Z3 \cite{z3}, CVC4 \cite{DBLP:conf/cav/BarrettCDHJKRT11}, and Alt-Ergo \cite{DBLP:conf/synasc/ConchonIM13}.

While the real model provides theoretical groundwork, the float model reflects the practical reality of finite precision computations in computer-based systems. Rounding errors in the float model can lead to inaccuracies when evaluating system properties. Our approach mitigates the challenges of verifying properties in the float model, challenges that tools like Frama-C often struggle with due to the complexities introduced by floating-point arithmetic.
We address this by explicitly quantifying the floating-point errors in the Hoare triple, accounting for errors introduced by finite arithmetic computations and rounding.
By adjusting the invariant properties to account for these errors, and subsequently verifying the corrected properties in the real model, we navigate around the limitations of existing tools. This strategy ensures that the properties verified are valid in real-world systems, despite the underlying verification being performed under the idealized assumptions of the real model. Thus, it not only establishes the theoretical correctness of system properties under idealized conditions but also confirms their validity and robustness in the practical, finite-precision computational environments where these systems are deployed.

\subsubsection{Verification in the Float Model}
For clarity, we detail the approach for correcting invariant properties for the following simplified equation:
\begin{equation}
    z = Ax+Bd,\label{eq:z_equation}
\end{equation}
where $z\in\mathbb{R}^{n_z}$, $x\in\mathbb{R}^{n_x}$, and $d\in\mathbb{R}^{n_d}$. We assume that the $i^{\text{th}}$ elements, $x_i$ and $d_i$, of $x$ and $d$, respectively, are bounded by $\bar{x}_i$ and $\bar{d}_i$, i.e., $|x_i|\leq\bar{x}_i$ and $|d_i|\leq\bar{d}_i$. A bound $e_i$, which overapproximates the floating-point error associated with the $i^{\text{th}}$ element of $z$, can be computed for $i = 1, \ldots, n_z$. The computation of such bounds is already established in the literature, as presented in \cite{Goubault:2001:SAP:647170.718304}, using the standard model of floating-point arithmetic~\cite{jeannerod:hal-00934443}, and implemented in tools of some theorem provers, such as the ``EVA'' plugin in Frama-C \cite{buhler2017structuring}.
After determining the error bounds, we define $e_z = \max_{i=1,\ldots,n_z} |e_i|$. The objective is to demonstrate that $\fl{z}$, the floating-point representation of $z$, falls within the ellipsoid $\mathcal{E}_X$ for a given positive definite matrix $X \in \mathbb{S}_{++}^{n_z}$. As established in \cite{NFM2023},
\begin{equation*}
    z \in \mathcal{E}_{\tilde{X}} \Rightarrow \fl{z} \in \mathcal{E}_X,
\end{equation*}
where $\tilde{X} =X/\alpha_z$, and $\alpha_z = 1-n_z e_z \lambda_{\text{max}}(X)\left( 2(\lambda_{\text{min}}(X))^{-\frac{1}{2}} + n_z e_z \right)$, with $\lambda_{\text{max}}(X)$ and $\lambda_{\text{min}}(X)$ being the maximum and minimum eigenvalues of $X$, respectively.
Then, to verify the property defined by the Hoare triple
\begin{equation*}
\{\text{Pre}(x) \land \text{Pre}(d)\} \; \texttt{compute$\_$z} \; \{\fl{z} \in \mathcal{E}_X\}
\end{equation*}
in the float model, where Pre$(x)$ and Pre$(d)$ are preconditions that impose the elementwise bounds of $x$ and $d$, respectively, and \texttt{compute\_z} is a function that computes $z$ as per (\ref{eq:z_equation}), it suffices to verify the following corrected property in the real model:
\begin{equation*}
\{\text{Pre}(x) \land \text{Pre}(d)\} \; \texttt{compute$\_$z} \; \{z \in \mathcal{E}_{\tilde{X}}\}.
\end{equation*}
Notably, $z \in \mathcal{E}_{\tilde{X}} \Rightarrow z \in \mathcal{E}_X$; therefore, if one of the preconditions is $z \in \mathcal{E}_X$, then verifying the postcondition $z \in \mathcal{E}_{\tilde{X}}$ implies this precondition and ensures it continues to be satisfied in subsequent iterations, thus establishing an invariant.

To apply this approach to the verification of the state and output invariant properties of system $H$ in the float model, we refer to (\ref{eq:augmentedsystem}). The elementwise bounds of $d(k)$ are derived from the precondition $d(k) \in \mathcal{D}$. Similarly, the elementwise bounds of $x_H$ and $y$ are obtained by projecting $\mathcal{E}_P$ and $\mathcal{E}_Q$, respectively, onto their corresponding subsets.
For the elements of $\vartheta$, the bounds $\bar{\vartheta}_i$, for $i=1,\ldots,n_\vartheta$, are computed using the bounds of $x_H$, $y$, and $d$, as well as the first and third equations in (\ref{eq:augmentedsystem}) and the triangle inequality, as follows:
    \[
    \bar{\vartheta}_i = \min_{\substack{1 \leq j \leq n_H \\ 1 \leq l \leq n_y}}{\left(\beta_{ji},\gamma_{li}\right)},
\]
where $X^{(ij)}$ denotes the element located at the $i^{\text{th}}$ row and $j^{\text{th}}$ column of the matrix $X$, $|x|$ corresponds to the absolute value of $x$, and
\begin{align*}
    \beta_{ji} &= \frac{1}{\left|B_{H_1}^{(ji)}\right|}\left(\sum_{k=1}^{n_H}\left(1+\left|A_{H}^{(jk)}\right|\right)\bar{x}_k+\sum_{k=1}^{n_d}\left|B_{H_2}^{(jk)}\right|\bar{d}_k\right),\\
    \gamma_{li} &= \frac{1}{\left|D_{H_{21}}^{(li)}\right|}\left(\sum_{k=1}^{n_H}\left|C_{H_2}^{(lk)}\right|\bar{x}_k+\sum_{k=1}^{n_d}\left|D_{H_{22}}^{(lk)}\right|\bar{d}_k+\sum_{k=1}^{n_y}\bar{y}_k\right).
\end{align*}

After establishing the elementwise bounds of $\vartheta$, $e_{x_H}$ and $e_y$ are computed similarly to $e_z$. Then, verifying the state and output invariant properties of $H$ in the float model corresponds to verifying the following corrected properties in the real model:
\begin{align*}
\{x_H(k) \in \mathcal{E}_P \land d(k) \in \mathcal{D} \land r(k)^\top S_x r(k) \geq 0\}\; &\texttt{stateUpdate}\; \{x_H(k+1) \in \mathcal{E}_{\tilde{P}}\},\\
\{x_H(k) \in \mathcal{E}_P \land d(k) \in \mathcal{D} \land r(k)^\top S_y r(k) \geq 0\}\;
&\texttt{outputUpdate}\;
\{y(k) \in \mathcal{E}_{\tilde{Q}}\},
\end{align*}
where $\tilde{P}=P/\alpha_{x_H}$, $\tilde{Q}=Q/\alpha_y$, and $\alpha_{x_H}$ and $\alpha_y$ are defined similarly to $\alpha_z$.

\section{Stateful Contracts: Verification of Invariant Systemic Properties Using Ghost Code}\label{sec:closedloopverification}
\begin{figure*}[t]
\centering
\includegraphics[scale=0.375]{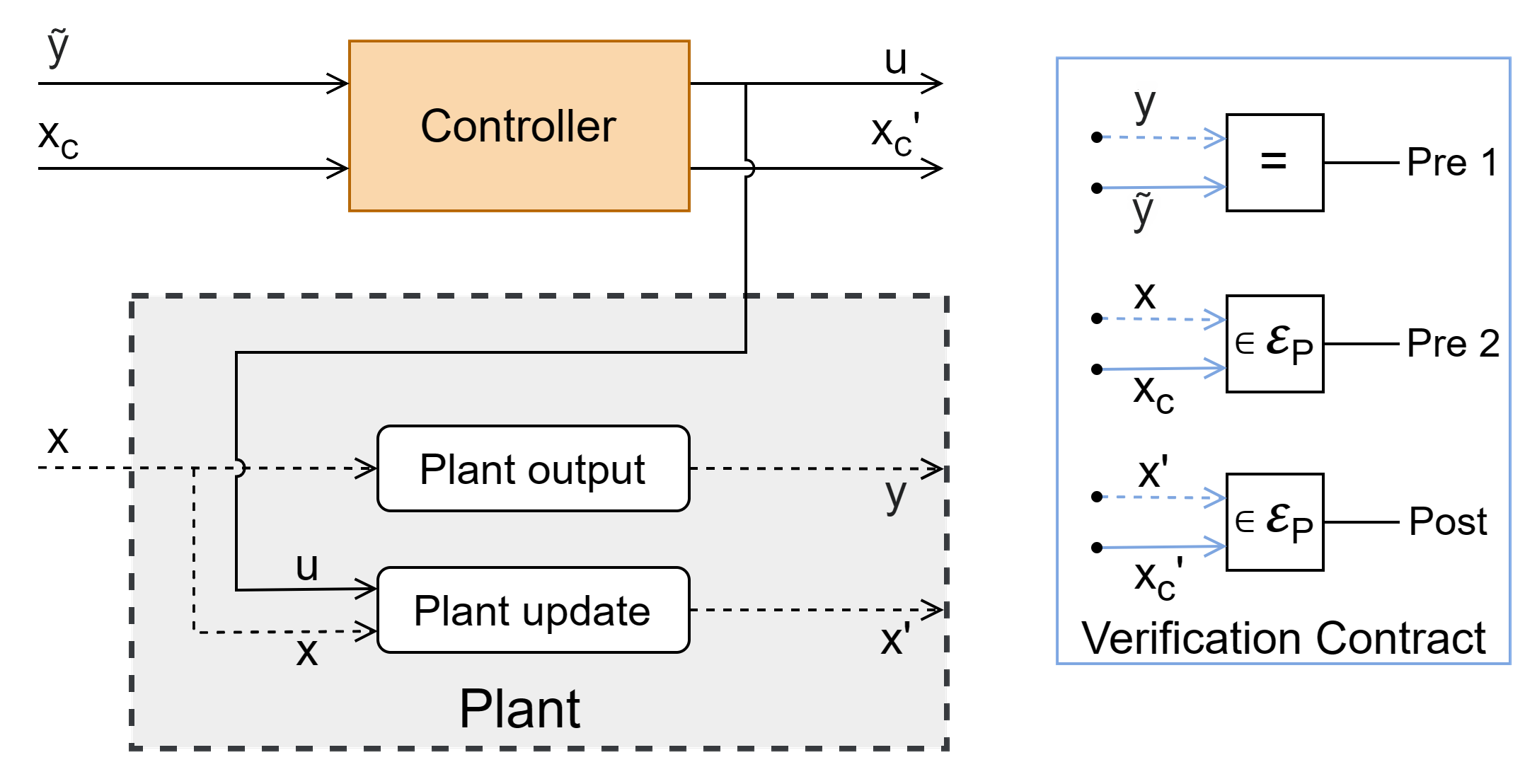}
\caption{Schematic of the verification process using ghost annotations, excluding exogenous inputs and uncertainties. The orange box represents the executable controller code; the dashed box corresponds to ghost code for the plant; dashed arrows indicate ghost variables; and variables with apostrophes are the updated states.}
\label{fig:ghost_verification_process}
\end{figure*}
Verifying the systemic properties guaranteed by a controller when interacting with plant dynamics, without altering the controller's executable source code, is crucial \cite{rtca2011airborne}. Some existing approaches attempt to verify these properties by expressing the entire closed-loop system (plant and controller) at the code level \cite{DBLP:conf/hybrid/RouxJG15}. However, this method has drawbacks: it modifies the controller code, thereby rendering it inappropriate for the verification of critical code \cite{DBLP:conf/hybrid/RouxJG15}. Consequently, the controller is not treated as an independent code entity interacting with the plant; instead, that interaction is implicit, which does not reflect real-world deployment where the controller interacts with the physical plant as a distinct entity. Furthermore, this method typically relies on simplified linear plant models \cite{wang2016formal} that fail to capture the inherent uncertainties due to inevitable modeling~errors.

In this work, we consider two types of controllers: LTI controllers and affine LPV controllers. In the case of LTI controllers, our approach enables the verification of systemic properties without modifying the executable controller code. We model the plant as an uncertain system that more accurately captures the dynamics of the physical plant and incorporate this model within code contracts as a stateful Hoare triple with the assistance of ghost memories. Specifically, ghost code \cite{filliatre2016spirit} is used to represent the uncertain system at the code level, allowing for accurate verification of systemic properties without affecting the controller’s executable code. See Fig. \ref{fig:ghost_verification_process} for an overview of this approach. In addition, an extension to this approach is provided  to tackle closed-loop systems consisting of an affine LPV controller and an uncertain plant, whereby the code-level verification of  systemic properties is conducted in both the real and float models and carried out in two steps: proving the systemic properties for  an augmented system, coupled with a quadratic constraint, that captures the behavior of the affine LPV controller and  verifying that this augmented system indeed overapproximates the unmodified controller code.

\subsection{LTI Controllers}\label{sec:lti_controller}
Consider the output feedback discrete-time LTI controller $K$ defined by the following state-space equations:
\begin{equation}
        x_c(k+1) = A_c x_c(k) + B_c y(k),\quad
        u(k) = C_c x_c(k) + D_c y(k),
 \label{eq:controller_equations}
\end{equation}
where the controller's state vector $x_c(k) \in \mathbb{R}^{n_c}$ and the control input vector $u(k) \in \mathbb{R}^{n_u}$ for all $k \geq 0$.
This controller is designed for an uncertain system $(G,\mathbf{\Delta})$, which is assumed to be well-posed and stabilizable, and  described by the following state-space equations:
\begin{equation*}
    \bmat{x_G(k+1)\\ \varphi(k) \\ y(k)} = \bmat{A_{ss}&A_{sp}&B_{1s}&B_{2s} \\ A_{ps}&A_{pp}&B_{1p}& B_{2p} \\ C_{s}&C_{p}&D&0} \bmat{x_G(k) \\ \vartheta(k) \\ d(k) \\ u(k)},\quad
    \vartheta = \Delta(\varphi),
\label{eq:system_with_control_input}
\end{equation*}
where $\Delta \in \mathbf{\Delta}$, $\mathbf{\Delta} \in \pwiqc(\Psi,S)$, $x_G(k) \in \mathbb{R}^{n_G}$, $y(k) \in \mathbb{R}^{n_y}$, $\vartheta(k) \in \mathbb{R}^{n_\vartheta}$, and $d(k) \in \mathbb{R}^{n_d}$ for all $k \geq 0$.

We then construct the augmented system $H$, with the augmented state $x_H = (x_G, x_\Psi)$. The equations of this system resemble those in (\ref{eq:augmentedsystem}), with the main difference being the additional terms corresponding to the control input $u$. Upon closing the loop between $H$ and $K$, the closed-loop system's state $x_{\text{cl}} = (x_H, x_c)$ is formed. Applying the result presented in Section \ref{sec:Preliminaries}, we can determine a state-invariant ellipsoid $\mathcal{E}_P$ for $x_{\text{cl}}$. The state invariant property of the closed-loop system classifies as a systemic property for the control program because this property is supposed to hold when the controller $K$ is connected to system $H$, which  captures the behavior of the uncertain system.

\subsubsection{Verification Using Ghost Code and Variables.}
The verification process uses ghost variables and ghost code, both declared with the \texttt{ghost} annotation, to model the uncertain system. Ghost code refers to non-executable segments within the source code, introduced to facilitate property verification without affecting the program's runtime behavior \cite{filliatre2016spirit}. Similarly, ghost variables simulate system states or variables that are crucial for verification while not interfering with the actual executable code.

The following Hoare triple demonstrates how ghost annotations are applied in verifying the behavior of a controller $K$ connected in feedback with an uncertain system modeled by the augmented system $H$:
\begin{align*}
\{\text{Pre}\} \; \texttt{controller} \; \{\text{Post}\},
\end{align*}

 \noindent where \texttt{controller} is the executable function that computes the control input and updates the controller's state according to (\ref{eq:controller_equations}),
\begin{align*}
    \text{Pre} &= \big(\left(x_H(k), x_c(k)\right) \in \mathcal{E}_P \land d(k) \in \mathcal{D}
     \land r(k)^\top S_x r(k) \geq 0\big),\\
    \text{Post} &= \left(x_H(k+1), x_c(k+1)\right) \in \mathcal{E}_P,
\end{align*}

 \noindent and the variables $(x_H,\vartheta,d)$ are declared as \texttt{ghost}. Note that $r$ is defined in terms of these ghost variables as indicated in  (\ref{eq:augmentedsystem}). In addition, $x_H$ and $y$ satisfy the state and output equations of the augmented system $H$, which are expressed in ghost code.
This Hoare triple represents the control system's state invariant property, a systemic property that can be directly verified in the real~model. Fig.~\ref{fig:sketch_acsl} presents the key elements of the expression of the closed-loop model as ACSL annotations.

\begin{figure}[t]
\begin{cacsl}
//@ predicate pwIQC_state(real r1,... , real r_nr) =
      Sx_11 * r1^2 + ... + Sx_nn * r_n^2 >= 0;
/*@ requires lower <= d <= upper;
    requires state_ellip(xp, *xc, 1);
    requires pwIQC_state(Hsys(xp, d, *xc, *u));
    requires y == plant_output(xp, d);
    ensures \let nxp = plant_update(xp, d, *u);
            state_ellip(nxp, *xc, 1); */
void controller(control_state *xc, control_output *u, control_input y) /*@ghost (plant_state xp, double d) */ { ... }
\end{cacsl}
\caption{Sketch of the  annotations used to attach closed-loop contracts to a controller code. The presented pseudo-code is a sketch that outlines the key elements of the contract. We refer the reader to the examples to see the valid syntax. Predicate \lstinline[style=acslstyle]!pwIQC_state! is called using the definition of the augmented system \lstinline[style=acslstyle]!Hsys(xp,d,*xc,*u)!. Similarly, \lstinline[style=acslstyle]!plant_output! and \lstinline[style=acslstyle]!plant_update! should define the dynamics of the augmented system. The presented contract requires to add an extra set of ghost variables (here \lstinline[style=acslstyle]!plant_state xp!) to the \lstinline[style=acslstyle]!controller! function prototype but does not modify the content of the controller function itself. \lstinline[style=acslstyle]!xp! only appears in the \lstinline[style=acslstyle]!requires! and \lstinline[style=acslstyle]!ensures!.}
\label{fig:sketch_acsl}
\end{figure}
\subsubsection*{Verification in the Float Model.}
The ghost variables are tied to the plant part of the closed-loop system and are not subject to floating-point errors, as they model the physical system. Thus, floating-point errors can be soundly restricted to the actual controller code, specifically affecting the controller state $x_c$ and the control input $u$.
The computation of the floating-point errors for $x_c$ requires knowledge of the bounds of the controller state variables and the measurements, which can be obtained by projecting $\mathcal{E}_P$ and $\mathcal{E}_Q$ onto the respective coordinates. Let $e_{x_c}$ denote the maximum floating-point error associated with the updated state $x_c$. To address these errors, we modify the postcondition as in Section \ref{sec:verification_process}.

Regarding the control input $u$, let $e_u$ denote the maximum floating-point error associated with computing $u$, and let $u_i^{\text{real}}$ correspond to the value of $u_i$ computed in the real model. The errors in the control input variables affect the updated state $x_H$. To account for these floating-point errors, we evaluate $x_H$ for all possible control inputs and their permissible perturbations due to floating-point errors. If the values of the updated state $x_H$ continue to satisfy the desired property, then the control input $u$ is deemed robust against floating-point errors.
The floating-point representation $\fl{u_i}$ of each control input variable $u_i$~satisfies
$u_i^\text{real} - e_u \leq \fl{u_i} \leq u_i^\text{real} + e_u$, and so the expression $u_i^\text{real} + l_i e_u$ for all $l_i \in [-1,1]$ captures all possible values of $u_i$ within the float model.

The following modified postcondition ensures that, for every allowable perturbation in the control input, the  next state stays within the modified ellipsoid~$\mathcal{E}_{\tilde{P}}$:
\[
    \text{Post} =\! \bigg(\big(\forall\, |l_i| \leq 1, u_i(k) = u_i^\text{real}(k) + l_i e_u,
    i = 1, \ldots, n_u\big) \!\implies\! x_\text{cl}(k+1) \in \mathcal{E}_{\tilde{P}}\bigg),
\]
where $\tilde{P} = P / \alpha$ and $\alpha = 1 - n_\text{cl} e_{x_c} \lambda_{\text{max}}(P)\left(2 \left(\lambda_{\text{min}}(P)\right)^{-\frac{1}{2}} + n_\text{cl} e_{x_c}\right)$, with $n_\text{cl}=n_H+n_c$, and $x_\text{cl}=(x_H,x_c)$.

\subsection{Affine LPV Controllers}\label{sec:affine_lpv_controller}
Consider an output feedback discrete-time affine LPV controller $\tilde{K}$ as defined in (\ref{eq:controller_equations}), where the state-space matrices $A_c(\delta(k))$, $B_c(\delta(k))$, $C_c(\delta(k))$,  $D_c(\delta(k))$ have an affine dependence on the scheduling parameters $\delta(k)=(\delta_1(k),\ldots,\delta_{n_\delta}(k))$. While it is feasible to apply the verification approach used for the LTI controller in Section \ref{sec:lti_controller}  directly to an affine LPV controller, difficulties may arise in successfully discharging proofs of systemic properties, particularly when the state-space equations of the controller involve multiple terms dependent on scheduling parameters.
To address this, our approach involves modeling the affine LPV controller as an augmented system $\tilde{H}$. We then formally verify that $\tilde{H}$ is a sound representation of $\tilde{K}$ and that the systemic properties for the closed-loop system comprising $\tilde{H}$ and the plant model $H$ hold. Since $\tilde{H}$ is shown to be a sound model of $\tilde{K}$, we can consequently infer that the systemic properties verified for the augmented system also hold for the executable affine LPV controller code.

\subsubsection{Affine LPV Controller Model Verification.}
Assuming one scheduling parameter for simplicity, i.e., $n_{\delta}=1$, the affine LPV controller $\tilde{K}$ can be equivalently expressed in LFT form with  uncertainty set $\mathbf{\Delta}_\mathrm{SLTV}$, which satisfies the pointwise IQC defined by $(\Psi_{\mathrm{SLTV}}, S_{\mathrm{SLTV}})$, as shown in Section \ref{sec:IQC_Framework}. To distinguish the $\varphi$, $\vartheta$, and $r$ variables of the controller from those of the plant model, we denote the controller variables with a tilde. Accordingly, we can construct an augmented system $\tilde{H}$, which, along with the pointwise IQC condition $\tilde{r}(k)^\top S_\mathrm{SLTV}\tilde{r}(k)\geq0$, models the affine LPV controller.
However, we need to formally verify that $\tilde{H}$ is a sound model of the executable controller code. Specifically, this verification requires showing that for any given current state $x_c(k)$, output measurement $y(k)$, and scheduling parameter $\delta(k)$, it is possible to construct $\tilde{\varphi}(k)$ and $\tilde{\vartheta}(k)$ such that the updated state and control input of $\tilde{H}$ match those of the executable controller code, while also satisfying the IQC condition.

To this end,  we verify at the code level that, at any time instant $k$,
\begin{gather*}
\forall x_c(k) \in \mathbb{R}^{n_c}, \, y(k) \in \mathbb{R}^{n_y}, \, |\delta(k)| \leq \alpha, \, \exists \, \tilde{\varphi}(k) \in \mathbb{R}^{n_{\tilde{\varphi}}} \text{ and } \tilde{\vartheta}(k) = \delta(k) \tilde{\varphi}(k) \\
\text{such that } x_{\tilde{H}}(k+1) = x_c(k+1), \, u_{\tilde{H}}(k) = u(k), \text{ and } \tilde{r}(k)^\top S_\mathrm{SLTV} \tilde{r}(k) \geq 0.
\end{gather*}
In essence, this verification confirms that the reachable states and control inputs of $\tilde{K}$ are contained within those of $\tilde{H}$, and that the constructed variables $\tilde{\varphi}(k)$ and $\tilde{\vartheta}(k)$ satisfy the pointwise IQC condition. In the case of  LPV systems, as shown in Section \ref{sec:IQC_Framework}, the pointwise IQC condition takes the form $\tilde{r}(k)^\top S_\mathrm{SLTV} \tilde{r}(k) = (\alpha^2 - \delta(k)^2)\tilde{\varphi}(k)^\top X \tilde{\varphi}(k) \geq 0$ for a given matrix $X \succeq 0$. Consequently, the verification of this property can be implied by verifying the more general property $(\alpha^2 - \delta(k)^2) z^\top X z \geq 0$ for all $z \in \mathbb{R}^{n_{\tilde{\varphi}}}$ and any $X \succeq 0$ since the considered values of $z$ would include those of the constructed $\tilde{\varphi}(k)$.

\begin{remark}\label{rmk:psd_matrix}
Verifying that an $n\times n$ symmetric matrix $X$ is positive semidefinite can be done in several ways. For instance, it can be formally proven by verifying that $z^\top X z \geq 0$ holds for all vectors $z \in \mathbb{R}^n$ using a tool like KeYmaera X \cite{fulton2015keymaera}. Alternatively, the verification can be empirically carried out by computing the eigenvalues of $X$ and ensuring that, within an acceptable margin, these eigenvalues are nonnegative. Once verified, this positive semidefiniteness property can then be passed as an axiom to tools like Frama-C, allowing them to assume it as a given fact for further analysis.
\end{remark}

\subsubsection{Verification of Systemic Properties in the Float Model.}
Upon closing the loop between $H$ and $\tilde{H}$, the closed-loop system’s state $x_{\text{cl}} = (x_H, x_{\tilde{H}})$ is formed. Applying the results from Section \ref{sec:Preliminaries}, we determine the state-invariant and output-bounding ellipsoids, $\mathcal{E}_P$ and $\mathcal{E}_Q$, respectively. We then follow similar steps to those presented in Section \ref{sec:lti_controller}, with the key difference being that $H$ is modeled using ghost code while $\tilde{H}$ is implemented as executable code. Since we have formally established that $\tilde{H}$ is a sound model for the executable controller code in the real model, verifying the systemic property for this closed-loop system implies that it holds for the controller’s executable code as well. Now, we extend this verification to the float model, as the executable code operates with finite precision in practice.

To carry out this task, we compute the floating-point errors, $e_{x_{\tilde{H}}}$ and $e_{u_{\tilde{H}}}$, associated with the state and control input of $\tilde{H}$, respectively, using bounds obtainable from projecting $\mathcal{E}_P$ and $\mathcal{E}_Q$ onto the appropriate planes. Similarly, we use these projected ellipsoid bounds to calculate the floating-point errors for the state and control input of the executable controller code, denoted as $e_{x_\mathrm{exe}}$ and $e_{u_\mathrm{exe}}$.
To ensure that the augmented system $\tilde{H}$ captures the floating-point behavior of the executable controller code, we define the effective error bounds as $e_{x_c} = \max(e_{x_{\tilde{H}}}, e_{x_\mathrm{exe}})$ and $e_u = \max(e_{u_{\tilde{H}}}, e_{u_\mathrm{exe}})$. Finally, we modify the postcondition used in Section \ref{sec:lti_controller} by incorporating these combined error bounds, resulting in the adjusted ellipsoid $\mathcal{E}_{\tilde{P}}$ as defined in Section \ref{sec:lti_controller}. Here, $e_{x_c}$ and $e_u$ are defined as the maximum error bounds to ensure that the verification accounts for the largest possible floating-point deviations across both the executable code and the augmented system. Verifying the modified systemic property in the real model thus ensures the validity of the original system property for the executable code in the float model as well.

\section{Examples}\label{sec:examples}
We verify the state and output invariant properties of UAS and AUV
controllers following the approach  in Section \ref{sec:VerificationOfInvariants}, and the systemic properties of an LTI $\mathcal{H}_\infty$  controller  for a  hovercraft, as proposed in Section \ref{sec:lti_controller}, in both the real and float models.
In the hovercraft case, the controller code remains untouched; all specifications and aspects related to the uncertain plant model are described in the ghost code. In addition, the systemic properties of an affine LPV controller designed for the two-mass rotational system from \cite{Farhood2008,Farhood2023CDC} are verified in both the real and float models following the approach in Section \ref{sec:affine_lpv_controller}. The annotated C scripts for all the examples  are available in the \href{https://github.com/ploc/submission_systemic_prop_IQC}{GitHub repository}.

\subsection{UAS Controller}\label{sec:UASexample}
We consider the robustly stable LPV path-following controller for a small, fixed-wing UAS designed in \cite{Deva2017}. The uncertainty set $\mathbf{\Delta}$ in this controller consists of static LTV uncertainty operators $\Delta$ that correspond to the multiplication in the time domain by a bounded, time-varying scalar $\delta(k)$. In other words, $\vartheta=\Delta(\varphi)$ corresponds to $\vartheta(k)=\delta(k)\varphi(k)$, where $|\delta(k)|\leq \bar{\delta}$ for all time instants $k$. This class of uncertainty operators admits a pointwise IQC characterization $(\Psi,S)$ \cite{veenman2016}; thus, the properties of this uncertain system can be verified by following the steps discussed in Section \ref{sec:VerificationOfInvariants}. The scheduling parameter $\delta$ is the inverse of the radius of curvature of the designated path to follow and $\bar{\delta}=1$. The controller has $n_G=16$ state variables, $n_y=4$ outputs, and $n_d=10$ inputs, and its state-space matrix-valued functions have polynomial dependence on the scheduling parameter $\delta$. The inputs and outputs of the controller are the measurements (specifically, the errors between the actual measurements and their reference values) and the control excursions (i.e., the deviations of the control commands from their reference values), respectively. The input bounds are determined based on flight-test data. The LTI augmented system $H$ is constructed based on (\ref{eq:augmentedsystem}), where $n_\vartheta=7$ and $n_r=14$. The state-invariant and output-bounding ellipsoids $\mathcal{E}_P$ and $\mathcal{E}_Q$, respectively, are computed along with $S_x$ and $S_y$ for pointwise polytopic input constraints using the results in Section \ref{sec:invariant_theorems}. To verify the state and output invariant properties, we write the annotated C script corresponding to $H$ following the template detailed in Appendix \ref{sec:frama-c_template}. In the float model, the floating-point errors are computed after finding the bounds of the elements of $\vartheta$, as discussed in Section \ref{sec:verification_process}. After executing the verification command, we obtain the  results
\begin{verbatim}
[wp] 24 goals scheduled
[wp] Proved goals:   24 / 24
  Qed:              20  (10ms-48ms-346ms)
  Alt-Ergo-Poly :    4  (17.8s-1'5s) (15667)
\end{verbatim}
These results confirm that the state and output invariant properties of $H$ are verified using the Alt-Ergo-Poly solver~\cite{DBLP:conf/tacas/RouxIC18}, while the other proof obligations are proven using WP's built-in solver Qed \cite{WPmanual}. Hence, we conclude that these state and output invariant properties are also verified for the UAS controller.

\subsection{AUV Controller}\label{sec:AUVexample}\vspace{-1mm}
We consider the robustly stable LPV path-following controller designed for the AUV built by the Center for Marine Autonomy and Robotics at Virginia Tech.
It is a tail controlled AUV equipped with four rear-mounted fins arranged in a T-tail configuration and a 3-bladed propeller in a pusher configuration; see \cite{njaka2022guide} for the details of the AUV model. The AUV operates at a fixed RPM of 1500, and thus the control inputs are the four fin angles. The path-following AUV controller is designed using a similar approach to the ones provided in \cite{Fry2019,muniraj2019lpv}. The uncertainty operator and the scheduling parameter $\delta$ with its bound $\bar{\delta}=1$ are defined similarly to those of the UAS controller.
The controller has $n_G=12$ states, $n_y=4$ outputs, and $n_d=12$ inputs. These inputs and outputs correspond to the measurement and control excursions, respectively. Bounds for these inputs are set based on field trials. The LTI augmented system $H$ is constructed based on (\ref{eq:augmentedsystem}), where $n_\vartheta=14$ and $n_r=28$. Based on the results in Section \ref{sec:invariant_theorems}, we compute the state-invariant and output-bounding ellipsoids $\mathcal{E}_P$ and $\mathcal{E}_Q$, along with $S_{x}$ and $S_{y}$, subject to pointwise polytopic input constraints. After writing of the annotated C script for $H$, we obtain the following verification results:
\begin{verbatim}
[wp] 24 goals scheduled
[wp] Proved goals:   24 / 24
  Qed:              20  (10ms-48ms-346ms)
  Alt-Ergo-Poly :    4  (17.8s-1'5s) (15667)
\end{verbatim}
These results confirm that the state and output invariant properties of $H$ are verified. Just as with the UAS controller, the Alt-Ergo-Poly solver verifies these properties, and WP's native solver Qed verifies the remaining proof obligations. Therefore, the desired invariant properties are verified for the AUV controller.

\subsection{Four-Thruster Hovercraft}\label{sec:hovercraftexample}
We consider the four-thruster hovercraft system from \cite{GRYMIN2014214}, which operates in a 2D plane. The system comprises $n_G=6$ state variables, specifically the positions $x$ and $y$, the heading angle $\theta$, and the velocities $\dot{x}$, $\dot{y}$, and $\dot{\theta}$. The hovercraft has $n_u=4$ control inputs, designating the thrusts generated by the four thrusters.
The nonlinear equations of motion describing the hovercraft dynamics are provided in \cite{GRYMIN2014214}. A continuous-time LPV model that closely captures the behavior of the nonlinear system in some desired envelope about a straight-path reference trajectory is derived following a similar approach to the one used in the example in \cite{FARHOOD20082108}. The chosen reference trajectory has a heading angle $\theta_{\text{ref}}=0$ and a linear velocity $(\dot{x}_{\text{ref}},\dot{y}_{\text{ref}})=(1,0)\,\mathrm{m/s}$. The LPV model has a polynomial dependence on the scheduling parameter $\bar{\theta}=\theta-\theta_{\text{ref}}=\theta$ and constitutes a close approximation of the nonlinear system provided that the condition $|\bar{\theta}|\leq 30^\circ$ holds for all time, in which case the errors induced by truncating the higher order terms in the Taylor series expansions become insignificant. The system's output vector consists of the position and heading errors $\bar{x}$, $\bar{y}$, and $\bar{\theta}$.
We assume that the LPV model is subjected to $n_d=6$ exogenous disturbances, manifesting as forces in the $x$ and $y$ directions, torques, and measurement noise. The explicit bounds for these disturbances are given in the annotated script. For instance, the measurement noise bounds are $0.2\,\mathrm{m}$ for both $x$ and $y$, and $1.5^\circ$ for $\theta$.

The continuous-time LPV model is expressed in LFT form using the LFR toolbox \cite{LFRtoolbox} in MATLAB. The corresponding discrete-time LFT system is derived via zero-order hold sampling, using a sampling time of $T=0.02\,\mathrm{s}$. 
The uncertainty operator in this system is defined similarly to that in the UAS controller. The augmented system $H$ is then constructed, with $n_\vartheta=14$ and $n_r=28$.
Subsequently, an $\mathcal{H}_\infty$  controller is designed using MATLAB's \texttt{hinfsyn} function, which utilizes the results in \cite{HinfitySynthesis}. The controller aims to ensure that the hovercraft closely tracks the straight-path trajectory and the condition, $|\bar{\theta}(k)| \leq 30^\circ$, holds for all discrete instants $k$. If this threshold is exceeded, the LPV approximation of the nonlinear system would no longer be acceptable.
After closing the loop between $H$ and the controller, we compute the state-invariant ellipsoid $\mathcal{E}_P$ along with $S_x$ using the results in Section \ref{sec:invariant_theorems}. Projecting $\mathcal{E}_P$ onto the $\bar{\theta}$-plane yields an upper bound for $|\bar{\theta}(k)|$ equal to $0.2501\,\mathrm{rad}\approx 14.3^\circ<30^\circ$, confirming the validity of the LPV approximation.
Following the steps presented in Section \ref{sec:closedloopverification} and the template in Appendix \ref{sec:frama-c-ghost}, we write the annotated C script and run the verification command to obtain the following verification results:
\begin{verbatim}
[wp] 11 goals scheduled
[wp] Proved goals:   11 / 11
  Qed:               9  (7ms-16ms-54ms)
  Alt-Ergo-Poly :    2  (18.1s-21.4s) (4871)
\end{verbatim}
These results confirm the verification of the systemic property of this control program in both the real and float models.

\subsection{Two-Mass Rotational System}\label{sec:twomassexample}
We consider the two-mass rotational system from \cite{Farhood2008,Farhood2023CDC} consisting of two bodies with moments of inertia $J_1=1$ and $J_2=0.1$, a spring, and a damper. The system comprises $n_G=4$ state variables, which are the angular displacements and velocities of the two bodies, $\theta_1$, $\theta_2$, $\dot{\theta}_1$, and $\dot{\theta}_2$. The system has $n_u=1$ control input and $n_d=1$ exogenous input $d$, each corresponding to a torque applied to body 1. The spring's stiffness coefficient is uncertain and time-varying but measurable at each moment in time, specifically, $k_s(\delta(k))=0.075+0.025\delta(k)$, where $|\delta(k)|\leq1$ for all $k\geq0$, and the damping coefficient is $b=0.004$. The equations of motion, given in \cite{Farhood2008}, are discretized via zero-order hold sampling, with a sampling time of $T=0.1\,\mathrm{s}$. The resulting discrete-time system is an LPV system with affine dependence on the scheduling parameter $\delta$.
Choosing the performance and measurement outputs to be $z(k)=(\theta_2(k),u(k))$ and $y(k)=\theta_2(k)$, respectively, we design an affine LPV controller $K$ for the system following the approach developed in \cite{Farhood2023CDC}, where the matrix $\Gamma$, specifying the uncertain initial state, is chosen as $\Gamma=\bmat{0.05 & 0 & 0.05 & 0}^\top$. The affine LPV controller obtained has $n_c=4$ state variables. The controller is expressed in LFT form, and following model reduction, the number of scheduling parameter copies is $n_{\tilde{\varphi}}=n_{\tilde{\vartheta}}=5$.
Uncertainty operators in both the plant and controller LFT systems are defined similarly to those in the UAS controller. The augmented systems $H$ and $\tilde{H}$ are constructed, with $n_\vartheta=1$ and $n_{\tilde{\vartheta}}=5$. Next, we verify that $\tilde{H}$ is a sound model of the executable code of the affine LPV controller $K$ by following the steps discussed in Section \ref{sec:affine_lpv_controller}. Running the verification command, we obtain~the~results
\begin{verbatim}
[wp] 8 goals scheduled
[wp] Proved goals:    8 / 8
  Qed:             5  (2ms-8ms-27ms)
  Alt-Ergo :       1  (74ms) (313)
  Z3 4.8.7:        2  (20ms-60ms) (186217)
\end{verbatim}
In these results, one of the proved goals, discharged by Z3, corresponds to the state and control input reachability condition, while the goal discharged by Alt-Ergo pertains to the general condition that implies the pointwise IQC condition. In the verification annotations, an axiom asserting that a $5\times 5$ matrix $X$ is positive semidefinite, supporting the verification of the pointwise IQC condition, is used. Since these goals have been successfully verified, we affirm that $\tilde{H}$ is a sound model of the executable controller code.
We then close the loop between $H$ and $\tilde{H}$ and compute the state-invariant ellipsoid $\mathcal{E}_P$ along with $S_x$, using results from Section \ref{sec:invariant_theorems}. Following the steps in Section \ref{sec:closedloopverification} and using the template in Appendix \ref{sec:frama-c-ghost}, we write the annotated C script where $H$ is expressed using ghost code to model plant dynamics, while $\tilde{H}$, as a sound model of $K$, is implemented as executable code. Running the verification command yields the following results:
\begin{verbatim}
[wp] 7 goals scheduled
[wp] Proved goals:    7 / 7
  Qed:               5  (3ms-9ms-26ms)
  Alt-Ergo-Poly :    2  (559ms) (1303)
\end{verbatim}
These results confirm that the systemic property of this control program is verified in real and float models, using Alt-Ergo-Poly. Hence, we conclude that the systemic property holds for the executable code of the affine LPV controller~$K$.

\section{Conclusion}\label{sec:conclusion}
This paper demonstrates the formal verification of invariant properties for uncertain systems presented in LFT form by leveraging the IQC framework. The properties of interest are expressed using Hoare logic in the real model, and methods for verifying these properties in the float model are proposed. Furthermore, the paper outlines a procedure for verifying the systemic properties of control programs in both the real and float models using ghost code. The applicability of these approaches is demonstrated using tools like Frama-C, ACSL, WP, and Alt-Ergo-Poly. To validate their efficacy, the  verification techniques were implemented on four examples. The examples' scripts are available at the \href{https://github.com/ploc/submission_systemic_prop_IQC}{Github repository}.
In future work, we aim to extend our approaches to verify, at the code level, reachability analysis properties of uncertain systems. Another topic of interest is to improve the revalidation of numerical properties at the code level. The outcome of Alt-Ergo-Poly \cite{DBLP:conf/tacas/RouxIC18} in Frama-C is not always predictable even when the mathematical property has already been validated. Additional annotations could support such validation.

\bibliographystyle{splncs04}
\bibliography{biblio}

\appendix

\section{LFT Framework}\label{sec:LFT}
Discrete-time uncertain systems are commonly encountered in the field of control theory. They can be derived using system identification techniques or by discretizing continuous-time uncertain systems whose equations of motion are defined by ordinary differential equations. Various discretization methods can be utilized, such as the Euler method and the trapezoidal method \cite{butcher2016numerical}. Errors resulting from these discretization methods can be incorporated into the system as additional uncertainties. In this work, we focus on uncertainty operators that admit pointwise IQC characterizations. Henceforth, for simplicity, the uncertainty operators are assumed to be linear memoryless operators.

Consider the uncertain system $G_{\delta}$ defined by the  state-space equations
\begin{align*}
    x(k+1)&=A(\delta)x(k)+B(\delta)d(k),\\
    y(k)&=C(\delta)x(k)+D(\delta)d(k),
\end{align*}
where $k$ denotes discrete time and $x$, $y$, $d$ denote the state, output, and input signals, respectively. The system $G_{\delta}$ can represent a plant model or a controller, such as a gain-scheduled controller. In the context of feedback controllers,  $d$ and $y$ correspond to the feedback signal and the control signal,~respectively.

Assuming at most rational dependence on the uncertainties, the system $G_{\delta}$ can be equivalently expressed in an LFT form, i.e., as  the interconnection of an LTI nominal model $G$ and a structured uncertainty operator $\Delta$, as depicted in Fig. \ref{fig:LFTsystem}. This LFT system, $(G,\Delta)$, is described by the  state-space equations
\begin{align*}
x_G(k+1)&=A_{G}x_G(k)+B_{G_1}\vartheta(k)+B_{G_2}d(k),\\
\varphi(k)&=C_{G_1}x_G(k)+D_{G_{11}}\vartheta(k)+D_{G_{12}}d(k),\\
y(k)&=C_{G_2}x_G(k)+D_{G_{21}}\vartheta(k)+D_{G_{22}}d(k),\\
\vartheta&=\Delta(\varphi),
\end{align*}
where $x_G(k)\in\mathbb{R}^{n_G}$, $d(k)\in\mathbb{R}^{n_d}$, $y(k)\in\mathbb{R}^{n_y}$, $\varphi(k)\in\mathbb{R}^{n_\varphi}$, $\vartheta(k)\in\mathbb{R}^{n_\vartheta}$, and $\Delta:\ell_2^{n_\varphi}\rightarrow\ell_2^{n_\vartheta}$. In general, the operator $\Delta$ can be structured and expressed as $\Delta=\mathrm{blkdiag}(\Delta_1,\Delta_2,\ldots,\Delta_N)$, for some positive integer $N$, where the diagonal blocks $\Delta_i$ are operators corresponding to different types of uncertainties, i.e., $\Delta_i\in \mathbf{\Delta_i}$ for $i = 1,\ldots,N$. The signals $\varphi$ and $\vartheta$ reflect the impact of the uncertainty  $\Delta$ on the system dynamics.
As the uncertainty operators are assumed to be linear and memoryless, the state-space matrices of $G_{\delta}$ and $(G,\Delta)$ are related as follows:
\begin{align*}
    A(\delta)&=A_G+B_{G_1}\Delta(I-D_{G_{11}}\Delta)^{-1}C_{G_1},\\
    B(\delta)&=B_{G_2}+B_{G_1}\Delta(I-D_{G_{11}}\Delta)^{-1}D_{G_{12}},\\
    C(\delta)&=C_{G_2}+D_{G_{21}}\Delta(I-D_{G_{11}}\Delta)^{-1}C_{G_1},\\
    D(\delta)&=D_{G_{22}}+D_{G_{21}}\Delta(I-D_{G_{11}}\Delta)^{-1}D_{G_{12}}.
\end{align*}
Deriving the LFT reformulation $(G,\Delta)$ of $G_{\delta}$ can be done manually or using some currently available toolboxes, such as the SMAC toolbox \cite{biannic2021generalized}.

To illustrate how an uncertain system can be represented in LFT form, consider the following numerical example:
\begin{align*}
x(k+1) = \bmat{a_1 && a_2+a_3\delta(k)+a_4\delta(k)^2 \\ a_5 && a_6 + a_7\delta(k)}x(k) + \bmat{b_1 \\ b_2} d(k),
\end{align*}
where $a_i$, for $i=1,\ldots,7$, and $b_j$, for $j=1,2$, are constant scalars, and $\delta(k)$ is an uncertain parameter such that $|\delta(k)|\leq 1$. The LFT form of this system is defined by the following state-space matrices:
\begin{align*}
A_{G} &= \bmat{a_1 && a_2 \\ a_5 && a_6}, \hspace{-5mm}& B_{G_1}&=\bmat{a_3 && a_4 \\ a_7 && 0}, \hspace{-5mm}&B_{G_2}&=\bmat{b_1 \\ b_2},\\
C_{G_1} &= \bmat{0 && 1 \\ 0 && 0}, &D_{G_{11}} &= \bmat{0&&0\\1&&0}, &D_{G_{12}} &= \bmat{0 \\ 0},
\end{align*}
and $\Delta(k)=\delta(k)I_2$, i.e., $\Delta\in\mathbf{\Delta}_{\mathrm{SLTV}}$ with $\alpha=1$. The equivalence of these two systems can be easily verified.

\section{Templates for Verification of Invariant Properties}\label{sec:frama-c_template}
This appendix presents two templates for verifying the invariant properties of uncertain systems, modeled by the system $H$ defined in (\ref{eq:augmentedsystem}) and coupled with the condition (\ref{eq:pwIQCcondition}). First, a pseudocode template outlines the methodology for verifying state and output invariants. Then, an annotated Frama-C script with ACSL annotations is provided, demonstrating the practical application of this approach, complete with detailed formal specifications.

\begin{pseudocode}[caption={Pseudo-Code for Verifying the State and Output Invariants of Uncertain Systems.},label={lst:pseudocode_localproperties}]
Define Structures:
  - state with elements $x_1,\ldots,x_{n_H}$.
  - output with elements $y_1,\ldots,y_{n_y}$.

Define Predicates:
  - state_ellip($x$, $\lambda$): true if $x^\top Px\leq\lambda$.
  - output_ellip($y$, $\lambda$): true if $y^\top Qy\leq\lambda$.
  - pwIQC_state($r$): true if $r^\top S_xr\geq0$.
  - pwIQC_output($r$): true if $r^\top S_yr\geq0$.

Function updateState($x_H(k)$, $\vartheta(k)$, $d(k)$):
  Update Logic:
    Compute new state values based on $(\ref{eq:augmentedsystem})$.
  Preconditions:
    * $x_H(k)$ is valid and has no memory overlap with other structures.
    * $d(k)\in\mathcal{D}$.
    * pwIQC_state($r(k)$).
  Contract under real model assumptions:
    * Ensure state_ellip($x_H(k)$,$1$) $\Rightarrow$ state_ellip($x_H(k+1)$,$1$).
  Contract under float model assumptions:
    * Ensure state_ellip($x_H(k)$,$1$) $\Rightarrow$ state_ellip($x_H(k+1)$,$\alpha_{x_H}$).

Function updateOutput($x_H(k)$, $y(k)$, $\vartheta(k)$, $d(k)$):
  Update Logic:
    Compute new output values based on $(\ref{eq:augmentedsystem})$.
  Preconditions:
    * $x_H(k)$ and $y(k)$ are valid and have no memory overlap with other structures.
    * $d(k)\in\mathcal{D}$.
    * pwIQC_output($r(k)$).
  Contract under real model assumptions:
    * Ensure state_ellip($x_H(k)$,$1$) $\Rightarrow$ output_ellip($y(k)$,$1$).
  Contract under float model assumptions:
    * Ensure state_ellip($x_H(k)$,$1$) $\Rightarrow$ output_ellip($y(k)$,$\alpha_{y}$).
\end{pseudocode}

In ACSL, annotations are introduced using \texttt{//@} for single-line specifications or \texttt{/*@ ... */} for block specifications. Here is a brief explanation of key ACSL annotations used in the script:
\begin{itemize}
  \item \texttt{requires}: Specifies preconditions that must hold true before the function executes.
  \item \texttt{ensures}: Specifies postconditions that must hold true after the function executes.
  \item \texttt{\textbackslash valid}: Checks the validity of pointers, ensuring they point to allocated and accessible memory.
  \item \texttt{assigns}: Lists the variables or memory locations that the function may modify, constraining side effects to specified locations.
  \item \texttt{\textbackslash separated}: Ensures specified pointers or memory regions do not overlap, preventing unintended side effects.
  \item \texttt{behavior}: Used to define specific behavior contracts for a function under certain conditions.
  \item \texttt{assumes}: Declares assumptions that are considered true within the scope of a behavior (or contract).
  \item \texttt{\textbackslash old}: Refers to the value of a variable before the function's execution (used in postconditions).
\end{itemize}
Once the annotated C script is ready, we verify the state and output invariant properties using the WP plugin \cite{WPmanual} and the Alt-Ergo-Poly solver \cite{DBLP:conf/tacas/RouxIC18}. This verification is performed by executing the following command:
\begin{verbatim}
frama-c -wp -wp-model real -wp-prover
Alt-Ergo-Poly script_name.c
\end{verbatim}
The \texttt{-wp-model real} option directs the WP plugin to perform calculations in the real arithmetic model. Additional command options are available, such as \texttt{-wp-timeout} to set a solver timeout \cite{WPmanual}.

\begin{cacsl}
typedef struct { double x1,... , x_nH; } state;

typedef struct { double y1,... , y_ny; } output;

//@ predicate state_ellip(real x1,... , real x_nH, real lambda) = P_11 * x1 * x1 + 2 * P_12 * x1 * x2 + ... + P_nHnH * x_nH * x_nH <= lambda;

//@ predicate output_ellip(real y1,... , real y_ny, real lambda) = Q_11 * y1 * y1 + 2 * Q_12 * y1 * y2 + ... + Q_nyny * y_ny * y_ny <= lambda;

//@ predicate pwIQC_state(real r1,... , real r_nr) = Sx_11 * r1 * r1 + ... + Sx_nrnr * r_nr * r_nr >= 0;

//@ predicate pwIQC_output(real r1,... , real r_nr) = Sy_11 * r1 * r1 + ... + Sy_nrnr * r_nr * r_nr >= 0;

/*@
requires \valid(x);

requires \separated(&(x->x1),... ,&(x->x_nH));

// Replace r1, r2,..., r_nr with their expressions in terms of x, theta, and d based on the state-space equations of the augmented system H.
requires pwIQC_state(r1,... , r_nr);

assigns *x;

behavior polytope_input_real_model:
	 assumes -d1_bar <= d1 <= d1_bar;
	 .
	 .
	 .
	 assumes -d_nd_bar <= d_nd <= d_nd_bar;

	 ensures state_ellip(\old(x->x1),... , \old(x->x_nH), 1) ==> state_ellip(x->x1,... , x->x_nH, 1);

behavior polytope_input_float_model:
	 assumes -d1_bar <= d1 <= d1_bar;
	 .
	 .
	 .
	 assumes -d_nd_bar <= d_nd <= d_nd_bar;

// Replace alpha_x with its expression in Section 3.2
	 ensures state_ellip(\old(x->x1),... , \old(x->x_nH), 1) ==> state_ellip(x->x1,... , x->x_nH, alpha_x);
*/

void updateState(state *x, double theta1,... , double theta_ntheta, double d1,... , double d_nd) {
	 double pre_x1 = x->x1,... , pre_x_nH = x->x_nH;

// Update the state vector based on the state-space equations of the augmented system H.
	 x->x1 = ...;
	 .
	 .
	 .
	 x->x_nH = ...;
}

/*@
requires \valid(x);
requires \valid(y);

requires \separated(&(x->x1),... ,&(x->x_nH),&(y->y1),... ,&(y->y_ny));

// Replace r1, r2,..., r_nr with their expressions in terms of x, theta, and d based on the state-space equations of the augmented system H.
requires pwIQC_output(r1,... , r_nr);

assigns *y;

behavior polytope_input_real_model:
	 assumes -d1_bar <= d1 <= d1_bar;
	 .
	 .
	 .
	 assumes -d_nd_bar <= d_nd <= d_nd_bar;

	 ensures state_ellip(\old(x->x1),... , \old(x->x_nH), 1) ==> output_ellip(y->y1,... , y->y_ny, 1);

behavior polytope_input_float_model:
	 assumes -d1_bar <= d1 <= d1_bar;
	 .
	 .
	 .
	 assumes -d_nd_bar <= d_nd <= d_nd_bar;

// Replace alpha_y with its expression in Section 3.2
	 ensures state_ellip(\old(x->x1),... , \old(x->x_nH), 1) ==> output_ellip(y->y1,... , y->y_ny, alpha_y);
*/

void updateOutput(state *x, output *y, double theta1,... , double theta_ntheta, double d1,... , double d_nd) {
	 double pre_x1 = x->x1,... , pre_x_nH = x->x_nH;

// Compute the output vector based on the state-space equations of the augmented system H.
	 y->y1 = ...;
	 .
	 .
	 .
	 y->y_ny = ...;
}
\end{cacsl}

\section{Templates for Verification with Ghost Code}\label{sec:frama-c-ghost}
This appendix provides a pseudocode template and a corresponding Frama-C script with ACSL annotations for verifying systemic properties of LTI control programs using ghost code. The template illustrates how ghost variables and functions model the dynamics of the plant, enabling verification without modifying the executable code.

\begin{pseudocode}[caption={Pseudo-code Template for Verifying Systemic Properties of a Control Program Using Ghost Code Annotations}]
Define Structures:
  - Controller State with elements $x_1, \ldots, x_{n_c}$.
  - Control Input with elements $u_1, \ldots, u_{n_u}$.
  - Output Measurements with elements $y_1, \ldots, y_{n_y}$.

Define Predicates:
  - state_ellip($x$, $\lambda$): true if $x^\top P x \leq \lambda$.
  - pwIQC_state(r): true if $r^\top S_x r \geq 0$.

Define ghost code and variables:
  - $\texttt{ghost}$ $x_H$.
  - $\texttt{ghost}$ $\vartheta$.
  - $\texttt{ghost}$ $d$.
  - $\texttt{ghost}$ Function plantDynamics($x_H(k)$,$\vartheta(k)$,$d(k)$,$u(k)$) that computes $x_H(k+1)$ based on the dynamics of $H$.

Function controller($x_c(k)$, $u(k)$, $y(k)$):
  Update Logic:
    Compute new controller state and input values based on $(\ref{eq:controller_equations})$
  Preconditions:
    * $x_c(k)$ and $u(k)$ are valid and have no memory overlap with other structures.
    * $d(k)\in\mathcal{D}$.
    * pwIQC_state($r(k)$).
    * $y(k)$ satisfies the output equation of $H$
  Contract under real model assumptions:
    * Ensure state_ellip($\left(x_H(k),x_c(k)\right)$,$1$) $\Rightarrow$
             state_ellip($\left(x_H(k+1),x_c(k+1)\right)$,$1$),
             where $x_H(k+1)\leftarrow$ plantDynamics($x_H(k)$,$\vartheta(k)$,$d(k)$,$u(k)$)
  Contract under float model assumptions:
    * Ensure $\forall$ $i\in\{1,\ldots,n_u\}$, $-1\leq l_i\leq 1$,
             state_ellip($\left(x_H(k),x_c(k)\right)$,$1$) $\Rightarrow$
             state_ellip($\left(x_H(k+1),x_c(k+1)\right)$,$\alpha$),
             where $x_H(k+1)\leftarrow$ plantDynamics($x_H(k)$,$\vartheta(k)$,$d(k)$,$u(k)+l\cdot e_u$), with $l=[l_1,\ldots,l_{n_u}]^\top$.
\end{pseudocode}

In addition to the ACSL annotations defined in Appendix \ref{sec:frama-c_template}, the annotated script in this appendix uses the following annotations:
\begin{itemize}
    \item \texttt{ghost}: Declares ghost variables or functions that do not affect the actual execution of the program but are used for verification purposes.
    \item \texttt{\textbackslash let}: Defines temporary variables that hold specific values for use in an expression.
    \item \texttt{logic}: Defines a logic function, which is a function in ACSL used for calculations within annotations.
    \item \texttt{\textbackslash at}: Allows referencing the value of a variable at a specific program point.
\end{itemize}
The same command as in Appendix \ref{sec:frama-c_template} is executed to verify the systemic properties of the control program.
\begin{cacsl}
/* First, we define
   - the state_ellip predicate describing the state invariant ellipsoid
   - one logic function for each plant state that corresponds to the computation of that state.
    (The logic functions are not mandatory but will facilitate writing contracts.)
*/

//@ predicate state_ellip(real x1,..., real x_ncl, real lambda) = ...;

//@ predicate pwIQC_state(real r1,... , real r_nr) = Sx_11 * r1 * r1 + ... + Sx_nrnr * r_nr * r_nr >= 0;

// Logic Functions to compute the augmented states based on x(k+1) = A_H*x(k) + B_H1*theta(k) + B_H2*d(k) + B_H3*u(k)
// Computing x1
/*@
  logic real update_x1 (real pre_x1, ..., real pre_x_nH, real d1, ..., real d_nd, real theta1, ..., real theta_ntheta, real u1, ..., real u_nu) =
  (Equation)
 */

// Computing x_nH
/*@
  logic real update_x_nH (real pre_x1, ..., real pre_x_nH, real d1, ..., real d_nd, real theta1, ..., real theta_ntheta, real u1, ..., real u_nu) =
  (Equation)
 */

typedef struct { double x1, ..., x_nc; } state;

typedef struct {double u1, ..., u_nu;} control_input;

/*@
    requires \valid(xc) && \valid(u);
    requires -d1_bar <= d1 <= d1_bar;
    .
    .
    .
    requires -d_nd_bar <= d_nd <= d_nd_bar;
    requires \separated(&(xc->x1),...,&(xc->x_nc),&(u->u1),...,&(u->u_nu));
    requires state_ellip(x1, ..., x_nH, xc->x1, ..., xc->x_nc, 1);

    // Replace r1, r2,..., r_nr with their expressions in terms of x_cl, theta, d, and u based on the state-space equations of the augmented system H.
    requires pwIQC_state(r1,... , r_nr);

    // Require that the output vector satisfies the output equations
    requires y1 == ...;
    .
    .
    .
    requires y_ny == ...;
    assigns *xc, *u;

    // Real model:
    ensures \let nx1 = update_x1 (\at(x1, Pre), ..., \at(x_nH, Pre), d1, ..., d_nd, theta1, ..., theta_ntheta, u->u1, ..., u->u_nu);
    	.
    	.
    	.
        \let nx_nH = update_x_nH (\at(x1, Pre), ..., \at(x_nH, Pre), d1, ..., d_nd, theta1, ..., theta_ntheta, u->u1, ..., u->u_nu);
        state_ellip(nx1, ..., nx_nH, xc->x1, ..., xc->x_nc, 1);

	// Float model (replace alpha with its expression in Section 4.1):
    ensures \forall real l_1; ...; \forall real l_nu;
        \let nx1 = update_x1 (\at(x1, Pre), ..., \at(x_nH, Pre), d1, ..., d_nd, theta1, ..., theta_ntheta, u->u1 + l_1 * e_u, ..., u->u_nu + l_nu * e_u);
    	.
    	.
    	.
        \let nx_nH = update_x_nH (\at(x1, Pre), ..., \at(x_nH, Pre), d1, ..., d_nd, theta1, ..., theta_ntheta, u->u1 + l_1 * e_u, ..., u->u_nu + l_nu * e_u);
        -1 <= l_1 <= 1 ==> ... ==> -1 <= l_nu <= 1 ==>
        state_ellip(nx1, ..., nx_nH, xc->x1, ..., xc->x_nc, alpha);
*/
void controller(state *xc, control_input *u, double y1, ..., double y_ny) /*@ ghost (double x1, ..., double x_nH, double theta1, ..., double theta_ntheta, double d1, ..., double d_nd) */ {
    double pre_xc1 = xc->x1, ..., pre_xc_nc = xc->x_nc;

    // Compute control inputs
    u->u1 = ...;
    .
    .
    .
    u->u_nu = ...;

    // Update controller states
    xc->x1 = ...;
    .
    .
    .
    xc->x_nc = ...;
}
\end{cacsl}
\end{document}